\documentclass[aps,twocolumn,superscriptaddress,floatfix]{revtex4-1}

\usepackage{amsmath,amsthm,verbatim,amssymb,amsfonts,amscd, graphicx}
\usepackage{graphics}
\usepackage[latin1]{inputenc}
\usepackage{url}
\usepackage[colorlinks=true,
            linkcolor=red,
            citecolor=blue,
            urlcolor=blue]{hyperref}

\newcommand{\SRO}{Sr$_2$RuO$_4$}

\usepackage{bm}

\newcommand{\bk}{\mathbf{k}}
\newcommand{\bq}{\mathbf{q}}
\newcommand{\bp}{\mathbf{p}}
\newcommand{\bx}{\mathbf{x}}

\newcommand{\up}{\uparrow}
\newcommand{\down}{\downarrow}
\newcommand{\avg}[1]{\left\langle #1 \right\rangle}

\newcommand{\be}{\begin{equation}}
\newcommand{\ee}{\end{equation}}

\newcommand{\bea}{\begin{equation}\begin{aligned}}
\newcommand{\eea}{\end{aligned}\end{equation}}

\begin{document}

\title{Degeneracy between even- and odd-parity superconductivity in the quasi-1D Hubbard model and implications for \SRO{}}
\author{Thomas Scaffidi}
\affiliation{Department of Physics and Astronomy, University of California, Irvine, California 92697, USA}
\affiliation{Department of Physics, University of Toronto, Toronto, Ontario, M5S 1A7, Canada}

\begin{abstract}
Based on a weak coupling calculation, we show that an accidental degeneracy appears between even- and odd-parity superconductivity in the quasi-1D limit of the repulsive Hubbard model on the square lattice. We propose that this effect could be at play on the quasi-1D orbitals Ru $d_{zx}$ and $d_{zy}$ of \SRO{}, leading to a gap of the form $\Delta_\text{even} + i \Delta_\text{odd}$ which could help reconcile several experimental results.
 \end{abstract}

\maketitle

\section{Introduction}
The presence of multiple components in the superconducting order parameter (OP) can lead to a flurry of interesting phenomena, like the spontaneous breaking of time-reversal symmetry (TRS) and the appearance of topological edge states \cite{RevModPhys.47.331,volovik2003universe,RevModPhys.83.1057}.
Multi-component superconductivity can either be symmetry-imposed, corresponding to a multidimensional irreducible representation (irrep) of the point group, or it can be accidental, when two superconducting orders are accidentally close to degenerate.
The latter scenario, although somewhat undesirable since it often requires fine tuning, has been invoked for a variety of superconductors \cite{PhysRevB.87.144511,PhysRevLett.117.027001,2020arXiv200200016K} for which a multi-dimensional irrep is in apparent contradiction with certain experiments, or when such an irrep does not exist altogether.

This work is motivated in particular by \SRO{}, for which the nature of the superconducting order remains an open question even 25 years after its discovery \cite{MaenoEA94,RiceSigrist95,BASKARAN1996490,MackenzieMaeno03,MaenoEA12,Kallin_2009,Kallin12,MackenzieEA17}.
This material sounds like a perfect testbed to study unconventional superconductivity, since its phase above $T_c$ is a well-behaved, albeit renormalized, Fermi liquid, for which Fermi surfaces have been measured with extreme accuracy \cite{DamascelliEA00, BergmannEA00, BergmannEA03, TamaiEA18}.
However, the theoretical study of this material has been hampered by several complications, including the presence of multiple orbitals (the quasi-1D orbitals $d_{xz}$ and $d_{zy}$ and the quasi-2D orbital $d_{xy}$) and their coupling via spin-orbit interaction.
Despite the challenges, achieving a consistent match between theory and experiments for this material would be an important milestone, and could shed new light on a flurry of other unconventional superconductors.

The evidence for TRS breaking \cite{LukeEA98, XiaEA06,2020arXiv200108152G} and multi-component superconductivity \cite{lupien2002ultrasound,2020arXiv200206130G,2020arXiv200205916B} in \SRO{} would naturally point towards a $\vec{d} = (p_x + i p_y) \hat{z}$ state.
However, such a state is in contradiction with the drop of spin susceptibility observed recently in NMR \cite{PustogowEA19,IshidaEA19}.
Several other candidates have thus been proposed \cite{PhysRevLett.121.157002,RamiresSigrist19,RomerEA19,2019arXiv191209525G,HuangEA19,HuangNematicEA19,2020arXiv200200016K}.
In particular, accidental degeneracies between non-symmetry-related orders have been considered, like $d+ i g$ \cite{2020arXiv200200016K} or $s'+ i d$ \cite{RomerEA19}.
Nevertheless, there is at least one experimental fact which seems difficult to explain for any candidate order parameter: the absence of a specific heat anomaly \cite{2019arXiv190607597L} at the putative second transition under $[1,0,0]$ strain revealed by muSR \cite{2020arXiv200108152G}.

In this work, we propose another candidate for a combination of accidentally degenerate states with the potential to resolve several of these issues: states of the form $\Delta_e + i \Delta_o$, where $\Delta_e$ is even-parity and $\Delta_o$ is odd-parity.
This proposal is based on our solution of the small-$U$ Hubbard model on a square lattice in the quasi-1D limit.
We provide an analytical proof that this model exhibits an accidental degeneracy between even and odd-parity representations (as previously pointed out in Ref.~\cite{Raghu2EA10}).
Since the Ru $d_{zx}$ and $d_{zy}$ orbitals in \SRO{} have a strongly 1D character, our hypothesis is that this mechanism could be at play on these orbitals, leading to a mixed parity order parameter on them.
Remarkably, $\Delta_e$ and $\Delta_o$ have the same magnitude everywhere on the Fermi surface, leading to a parametrically small specific heat jump.
This mechanism therefore provides a microscopic justification for an accidental degeneracy, along with a justification for a parametrically small second specific heat jump.

In Section II, we provide an exact analytical solution for weak coupling superconductivity in the repulsive Hubbard model for a quasi-1D band on the square lattice.
We show that there is an accidental degeneracy between even and odd-parity superconducting orders across the entire spectrum, and that this degeneracy is robust to changes in the dispersion relation.
In Section III, we use a Ginzburg Landau analysis to study the possible combinations of even and odd-parity SC orders. 
We find that states of the type $\Delta_e + i \Delta_o$ are favored.
We then study two thermodynamic properties of these states: specific heat and spin susceptibility. 
In Section IV, we assume this mechanism is at play on the quasi-1D bands of \SRO{} and discuss the consequences for experiments.

\section{Weak coupling calculation}

\begin{figure*}[t!]
\center
  \includegraphics[width=6cm]{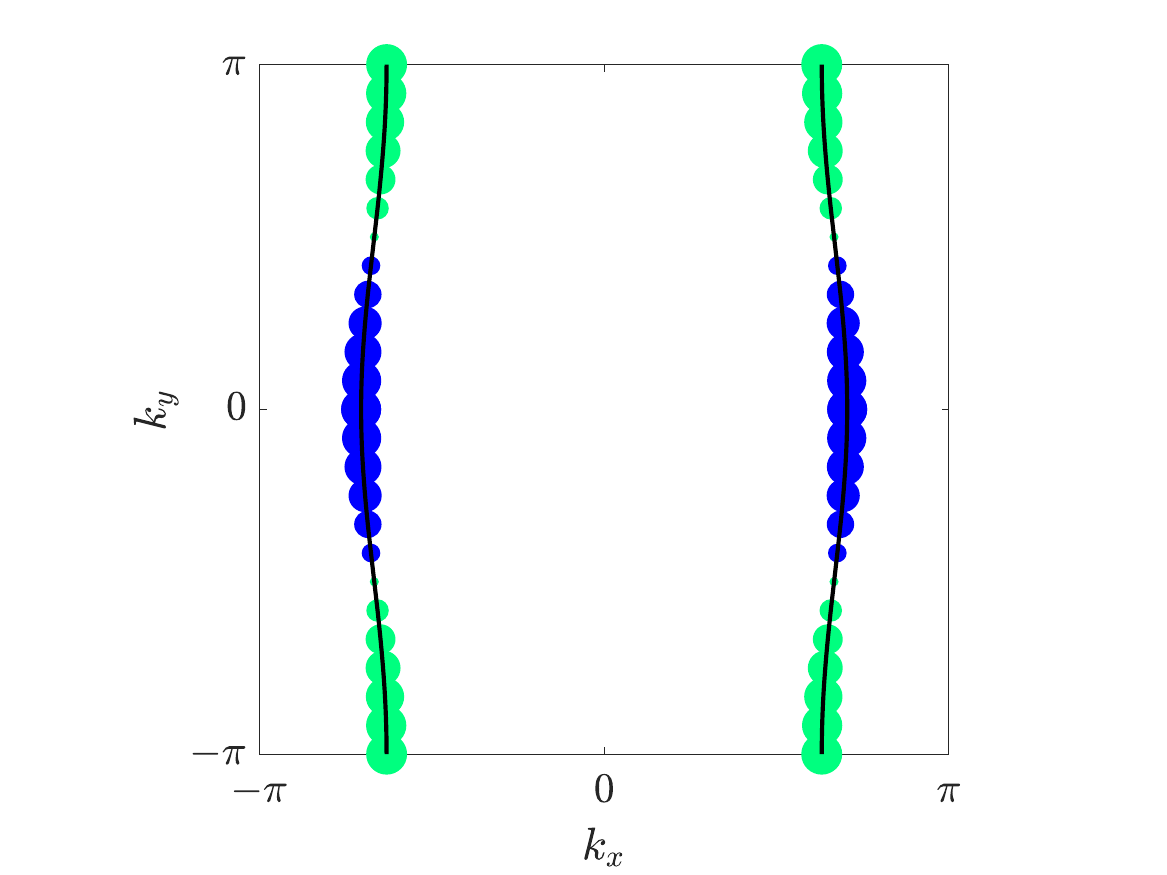}
   \includegraphics[width=6cm]{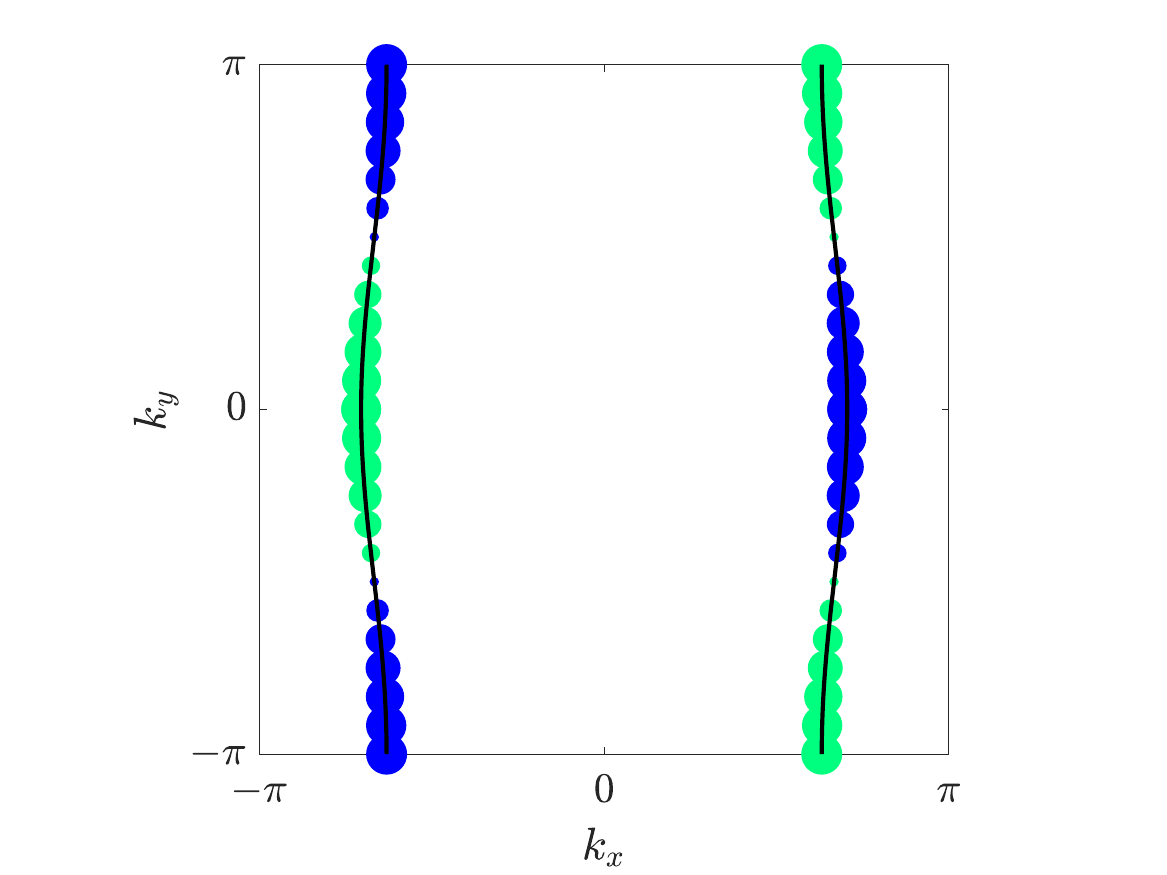}
  \caption{Nearly-degenerate dominant gap functions in the even (left) and odd (right) parity sector, evaluated at the Fermi surface, for $t_x=1, t_y=0.1,\mu=1$. The size of the dots gives the gap magnitude and the color gives the sign. The gap functions are well approximated by the analytical form of Eq.~\ref{GapForm} ($\Delta_{m=1,e}$ and $\Delta_{m=1,o}$) obtained in the limit of $t_y/t_x \rightarrow 0$.}
  \label{FigGaps}
\end{figure*}

We study a single orbital repulsive Hubbard model on a square lattice, with nearest-neighbor hoppings $t_x$ along the $x$ direction and $t_y$ along the $y$ direction.
The Hamiltonian reads
\bea
H = \sum_{\bk} \xi(\bk) (n_{\bk,\uparrow} + n_{\bk,\downarrow}) + \sum_{\bx} U n_{\bx,\uparrow}  n_{\bx,\downarrow}  
\eea
with a dispersion relation given by
\bea
\xi(k) = -2 t_x \cos(k_x) - 2 t_y \cos(k_y) - \mu.
\eea
We are interested in the quasi-1D limit: $t_y \ll t_x$.
In that limit, the Fermi surfaces are given by slightly corrugated vertical lines (see Fig \ref{FigGaps}):
\bea
k_F(k_y) = k_F + \frac{2 t_y \cos(k_y)}{v_F} + \mathcal{O}(t_y^2)
\eea
with $k_F = \mathrm{arccos}(- \mu / 2 t_x)$ and $v_F = 2 t_x \sin(k_F)$.

Following the standard weak coupling approach~\cite{KohnLuttinger65, BaranovEA92, KaganChubukov89, ChubukovEA92, BaranovChubukovEA92, ChubukovEA93, HironoEA02, Hlubina99, RaghuEA10, Raghu2EA10, WeejeeEA14,ScaffidiEA14, SimkovicEA16, Scaffidi2017, RoisingEA18}, valid in the limit $U/t \rightarrow 0$, we have to solve the following eigenvalue problem:
\bea
\frac1{(2\pi)^2} \int_{\text{FS}} \frac{d\hat{k}_2}{v(\hat{k}_2)} V(\hat{k}_1 - \hat{k}_2) \Delta(\hat{k}_2) = \lambda \Delta(\hat{k}_1)
\label{gapEq}
\eea
where the integral is over the Fermi surface, $V$ is the effective interaction in the Cooper channel, $v(\hat{k})$ is the norm of the Fermi velocity at momentum $\hat{k}$.
Each solution with negative eigenvalue $\lambda$ corresponds to a superconducting order with gap function $\Delta(\bk)$ and critical temperature $T_c \propto W e^{\frac1{\lambda}}$, with $W$ the bandwidth. The dominant order parameter has the most negative eigenvalue.

Since we are taking two limits ($U/t \rightarrow 0$ and $t_y / t_x \rightarrow 0$), it is important to specify the order in which they are taken.
We first take the weak coupling limit before taking the quasi-1D limit, which means that the system above $T_c$ behaves as a 2D Fermi liquid (as opposed to a Luttinger liquid if the other order of limits had been chosen).
This order of limits therefore allows us to use a weak coupling approach in a quasi-1D system, even though this approach is not valid in a strictly one-dimensional system \cite{giamarchi2004quantum}.
Note that the present model also differs from the case of small-$U$ multi-leg Hubbard ladders\cite{PhysRevB.53.12133}, since we work directly in the thermodynamic limit in both the $x$ and $y$ directions.

In a single orbital model, $V$ takes a simple form~\cite{Raghu2EA10}\footnote{Note that Ref.~\cite{Raghu2EA10} uses $V_e = U + U^2 \chi(\hat{k}_1 + \hat{k}_2)$ instead of $V_e = U + U^2 \chi(\hat{k}_1 - \hat{k}_2)$, but these two choices are equivalent since we only care about integrals of the type $\int d\hat{k}_2 V_e(\hat{k}_1,\hat{k}_2) \Delta(\hat{k}_2)$ with even gap functions ($\Delta(\hat{k}_2)=\Delta(-\hat{k}_2)$). It is easy to see that these integrals are equal for either choice of $V_e$. }:
\bea
V_e(\hat{k}_1 - \hat{k}_2) &= U + U^2 \chi(\hat{k}_1 - \hat{k}_2) \\
V_o(\hat{k}_1 - \hat{k}_2) &= - U^2 \chi(\hat{k}_1 - \hat{k}_2) 
\label{EffectiveInteraction}
\eea
in the even and odd-parity channel, respectively, and where $\chi(\bq)$ is the Lindhard susceptibility:
\bea
\chi(\bq) = \frac{-1}{(2 \pi)^2} \int d\bk \frac{n(\xi(\bk)) - n(\xi(\bk+\bq))}{\xi(\bk) -\xi(\bk +\bq)}
\eea
with $n(\xi)$ the Fermi-Dirac distribution.

\begin{figure*}
	\center
  \includegraphics[width=8cm]{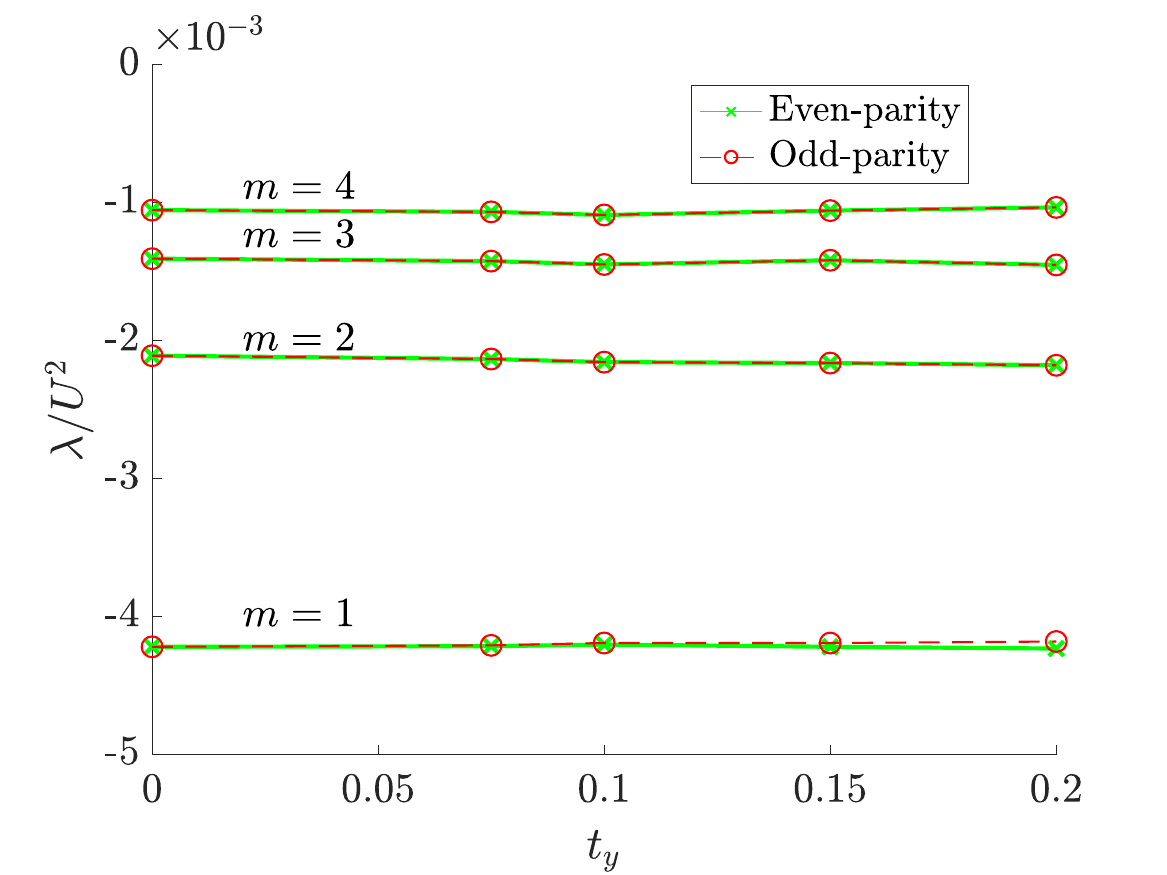}
    \includegraphics[width=8cm]{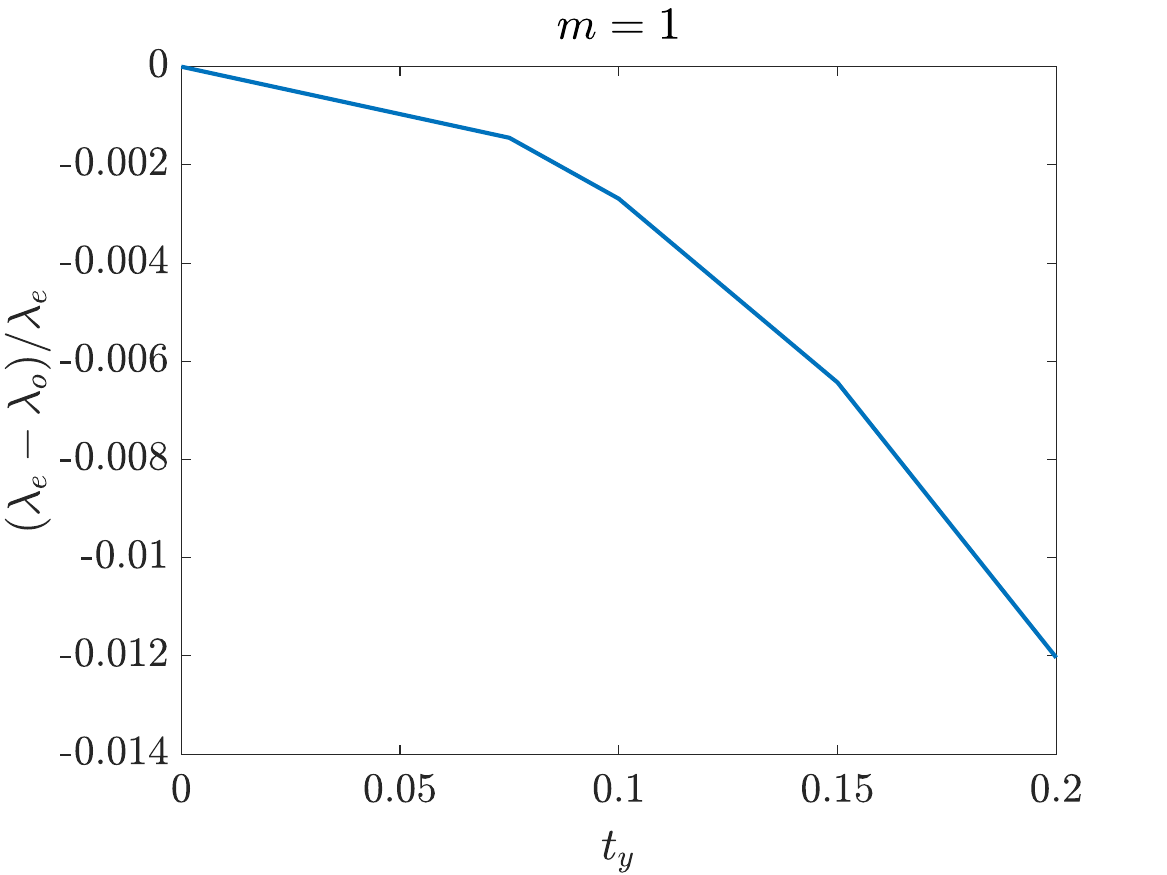}
  \caption{Left: Four dominant eigenvalues in the even and odd parity sectors, for $t_x=1,\mu=1$. The points at $t_y=0$ were obtained analytically from Eq.\ref{lambda_m}, whereas the points at $t_y>0$ were obtained by numerically solving Eq.~\ref{gapEq}. The splitting between even and odd-parity eigenvalues is so small that it is barely visible. Right: Splitting between the dominant (i.e. $m=1$) even and odd-parity eigenvalues, normalized by $\lambda_e$. The splitting increases quadratically with $t_y$ but remains extremely small for a wide range of $t_y$. 
  \label{Fig1}}
\end{figure*}

As explained in the Appendix, Eq.~\ref{gapEq} is analytically solvable in the limit of $t_y / t_x \rightarrow 0$, leading to the following negative eigenvalues:
\bea
\lambda_{m} = -  \frac{U^2}{2 (2\pi)^2 v_F^2} \frac1m
\label{lambda_m}
\eea
for $m=1,2,3,\dots$. The most negative eigenvalue thus corresponds to $m=1$.
Each eigenvalue is doubly degenerate, with an even and an odd-parity eigenvector. For odd $m$, these eigenvectors are given by
\bea
\Delta_{m,e} &= \cos(m k_y) \\
\Delta_{m,o} &=  \cos(m k_y) \mathrm{sign}(k_x).
\label{GapForm}
\eea
For even $m$, we find
\bea
\Delta_{m,e} &= \sin(m k_y)  \mathrm{sign}(k_x)\\
\Delta_{m,o} &= \sin(m k_y).
\eea

For each $m$, we therefore have two degenerate eigenvectors which are simply related by a sign change between the left and right branches of the Fermi surface.
The source of this degeneracy can be understood easily \cite{Raghu2EA10}.
In the quasi-1D limit, the almost perfect nesting of the Fermi surfaces leads to a strong peak in $\chi(\bq)$ for $q_x = \pm 2 k_F$. This means that the dominant type of scattering occurs between the two branches of the Fermi surface.
By flipping the relative sign of the gap on the two branches, one can therefore effectively flip the sign of the effective interaction.
This sign change exactly cancels out the sign difference for the $U^2$ term in the effective interaction between even and odd parity (see Eq.~\ref{EffectiveInteraction}) \footnote{The $U$ term in the singlet channel gives a zero contribution for any $m>0$.}.

Whereas the analytic results provided so far were obtained in the limit of $t_y / t_x \rightarrow 0$, we also studied numerically the case of small but finite $t_y/t_x$.
As shown in Fig.~\ref{Fig1}, the dependence on $t_y$ is extremely weak, and our analytic solution is therefore a good approximation for a broad range of $t_y/t_x$.
The main effect of a finite $t_y$ is to generate a small splitting between even and odd-parity states, which, for $m=1$, favors the even-parity state.
However, as shown in the right panel of Fig.~\ref{Fig1}, the splitting remains extremely small even for $t_y/t_x \simeq 0.1$, which is the range relevant for \SRO{}.
The effect of finite $t_y/t_x$ on eigenvectors is also small: they are still very well approximated by the simple cosine form given above even at $t_y/t_x = 0.1$.
We also checked that changing the chemical potential does not produce any qualitative changes to these results.

\section{Ginzburg-Landau analysis and thermodynamic properties}
In the previous section, we learned that the dominant superconducting orders in the quasi-1D Hubbard model are given by the two nearly-degenerate $m=1$ states:
\bea
\Delta_e \equiv \Delta_{m=1,e} &= \cos(k_y) \\
\Delta_o \equiv \Delta_{m=1,o} &= \cos(k_y)  \mathrm{sign}(k_x).
\label{Gaps}
\eea
In this section, we use a Ginzburg-Landau analysis to study the possible combinations of these two order parameters.

A combination of a singlet and triplet order parameter is non-unitary unless the relative phase between them is $\pm i$.
Complex combinations of singlet and triplet are therefore generically favored \cite{PhysRevLett.119.187003}.
We will thus consider the following order parameter:
\bea
\mathbf{\Delta}(\bk) \equiv \Delta_{\up\down}(\bk) = \psi_e \Delta_e(\bk) + i \psi_o \Delta_o(\bk)
\label{ComplexComb}
\eea
with $\psi_e$ and $\psi_o$ real parameters, leading to $|\Delta(\bk)|^2=|\psi_e \Delta_e(\bk)|^2 + |\psi_o \Delta_o(\bk)|^2$ \footnote{$\Delta_o$ should actually be described by a vector order parameter $\vec{d}$ describing the spin component of the Cooper pair\cite{Sigrist05}, but we are free to choose the orientation of $\vec{d}$ without loss of generality since the model is $SU(2)$-symmetric. We chose $\vec{d} \parallel \hat{z}$ in this section to simplify notations.}.
A typical GL free energy functional reads \footnote{We have omitted a term proportional to $\psi_e^4 - \psi_o^4$ for simplicity because it does not change the physics qualitatively.}
\bea
F = -a_e \psi_e^2 - a_o \psi_o^2 + b (\psi_e^2 + \psi_o^2)^2 + b' (\psi_e^2 - \psi_o^2)^2
\eea
with $a_e(T) \propto (T_{c,e} -T)$, $a_o(T) \propto (T_{c,o} -T)$, where $T_{c,e}$ and $T_{c,o}$ are the critical temperatures for each component when considered in isolation.

For $b'<0$, the system favors having only one component at a time, whereas for $b'>0$ the system favors a combination of the two.
We will see below that $b'$ is positive for the order parameters obtained in the previous section, so we will focus on that case.
 The small splitting between eigenvalues which slightly favors the even-parity order (see Fig.~\ref{Fig1}) translates into a small difference between the critical temperatures: $T_{c,e} = T_{c,o} + \delta$, with $\delta > 0$ small.
In this scenario, the $\psi_e$ component arises at the first transition $T_{c,e}$, and the $\psi_o$ component arises at a second transition $T^*$ given by
$
T^* = T_{c,o} - \delta \frac{b-b'}{2 b'}
$. We give a derivation of these results based on a Ginzburg-Landau analysis in Appendix \ref{App:GL}. An important approximation which was used in that analysis is that we assume the linear coefficients for $a_e(T) \propto (T_{c,e} -T)$ and $a_o(T) \propto (T_{c,o} -T)$ are equal.
 This assumption is justified by the fact that the two order parameters have essentially the same magnitude everywhere on the Fermi surface, as we now discuss.

Whereas the Ginzburg-Landau analysis presented so far is standard, 
what is unusual about $\Delta_e$ and $\Delta_o$ is that they 
have the same magnitude everywhere on the FS (see Fig.~\ref{FigGaps}):
\bea
|\Delta_e(\bk)|^2 \simeq |\Delta_o(\bk)|^2 \ \forall \ \bk \in \text{FS}.
\eea
This property is really unique since one usually considers combinations of OPs that gap out different parts of the Fermi surface (like $p_x + i p_y$ or $d_{x^2 - y^2} + i g_{(x^2 - y^2)xy}$).
The main consequence is that the parameter $b'$ is parametrically small (in $t_y/t_x$), in contrast to standard two-component order parameters for which it is of order one.
This can be deduced from the following microscopic formula~\cite{Frank:2016aa} for $b'$ :
\bea
\frac{b'}{B} &= \frac12 \left( \frac12 \avg{ |\Delta_e|^4 } + \frac12 \avg{ |\Delta_o|^4 } \right) - \frac12 \avg{ |\Delta_e|^2 |\Delta_o|^2 }
\label{bp}
\eea
where $B= \frac{7 \zeta(3)}{16 \pi^2 (k_B T_c)^2} \rho$ 
and where $\left\langle \dots \right\rangle$ is a Fermi surface average defined by
\bea
\left\langle f \right\rangle = \frac1{\rho} \frac1{(2 \pi)^D} \int_{FS} d\hat{k} \frac1{v(\hat{k})} f(\hat{k})
\eea
with $\rho$ the density of states at the Fermi level.
The difference in Eq. \ref{bp} is usually of order one (e.g. for $p_x + i p_y$ or $d_{x^2 - y^2} + i g_{(x^2 - y^2)xy}$), but in our case it is parametrically small.
In other words, a unique feature of the current scenario is that the small parameter $t_y/t_x$ leading to the near-degeneracy of critical temperatures also leads to a small $b'$ parameter.

\subsection{Specific heat}

An important consequence of a small $b'$ is that the jump in specific heat at the second transition $T^*$ is parametrically small.
As shown in Appendix \ref{App:CJump}, the ratio of specific heat jumps is given by 
\bea
\frac{\Delta C_{T^*}}{\Delta C_{T_c}} = \frac{\avg{|\Delta(\bk)|^2 Y_{T^*}(\bk)}}{\avg{|\Delta (\bk)|^2}} \frac{b'}{b}
\label{SpecHeat}
\eea
with
\bea
Y_{T}(\bk) &= \frac{1}{4} \int_{-\infty}^{\infty} dx \frac1{\cosh\left(\frac12  \sqrt{x^2 +\beta^2 |\mathbf{\Delta}(\bk)|^2} \right)^2} 
\eea
the $k$-dependent Yosida function and $\beta=1/k_B T$.
When deriving Eq.~\ref{SpecHeat}, we made the approximation that $|\Delta_e(\bk)|^2=|\Delta_o(\bk)|^2\equiv |\Delta(\bk)|^2$.

From Eq.~\ref{SpecHeat}, we learn that two separate effects can lead to a reduction of the second specific heat jump: the effect of the Yosida function, and the effect of a small $b'/ b$ ratio.
The first effect is always present for any two-component OP, and would act in the same way in this case~\cite{2020arXiv200200016K}.
However, this effect can only give a substantial reduction of the second specific heat jump if $T^*$ is much smaller than $T_{c,e}$.
On the other hand, the effect of $b' \ll b$ is unique to the current scenario, and naturally leads to a parametric difference between the two specific heat jumps.
As an illustration, for the numerical solution at $t_y/t_x=0.1$ obtained in the previous section, we find that $b'/b \sim 10^{-5}$.

\subsection{Spin susceptibility}
The even- and odd-parity components have of course different effects on the spin susceptibility since the former is a spin singlet and the latter is a spin triplet (We neglect spin-orbit coupling for the time being).
Taking advantage of the $SU(2)$ symmetry of the Hubbard model, we did not have to specify the orientation of $\vec{d}$ for the odd-parity, spin-triplet component in the previous discussion (see e.g. Ref.~\cite{Sigrist05} for a definition of $\vec{d}$).
It is however now necessary to specify it in order to discuss the spin susceptibility $\chi$.
Whereas the susceptibility of a spin singlet goes to zero for any orientation of the magnetic field $\vec{H}$, the situation is more complex when a spin triplet component is present.
When $\vec{H}$ is parallel to $\vec{d}$, the singlet and triplet component lead to the same decay of $\chi$, with zero residual spin susceptibility.
When $\vec{H}$ is perpendicular to $\vec{d}$, the spin susceptibility is given by (see Appendix \ref{App:SpinSus} for a derivation):
\bea
 \frac{\chi(T)}{\chi_N} =   \avg{Y_T(\bk)} +   \avg{  \frac{|\psi_o\Delta_o(\bk)|^2}{|\mathbf{\Delta}(\bk)|^2} \ (1-Y_T(\bk))}
 \label{Eq:MagSus}
\eea
with $\chi_N$ the normal state Pauli susceptibility. 
 At $T=0$, one finds $Y_T(\bk)=0$, leading to
\bea
 \frac{\chi(T=0)}{\chi_N} =      \avg{  \frac{|\psi_o\Delta_o(\bk)|^2}{|\mathbf{\Delta}(\bk)|^2}} \simeq \frac{|\psi_o|^2}{|\psi_e|^2+|\psi_o|^2}
 \eea
 where we made the approximation that $|\Delta_e(\bk)|^2 = |\Delta_o(\bk)|^2$ in the last step.
Assuming $|\psi_o|^2 \simeq |\psi_e|^2$ at $T=0$ (which is expected if the two critical temperatures are close to each other), this leads to a residual susceptibility of $\frac12$ for $\vec{d} \perp \vec{H}$.

\section{Application to Strontium Ruthenate}

As mentioned in the introduction, the main motivation behind this work is the study of superconductivity in \SRO{}.
The Hamiltonian studied above provides a good model for the quasi-1D Ru orbital $d_{zx}$ (and of course for $d_{zy}$ after a $\pi/2$ rotation) of \SRO{}, if it could be considered in isolation.
In this section, we will make the assumption that the above mechanism for accidental mixed-parity superconductivity is at play on each of these two orbitals, and we will analyze the consequences for experiments.
We should emphasize that this assumption is purely empirical: we do not claim to have a microscopic justification for neglecting the coupling between the two quasi-1D orbitals, and between the quasi-1D orbitals and the $d_{xy}$ orbital.

As thermodynamic measurements give evidence for a superconducting order of similar size on the three orbitals, we also need to make an assumption about the OP on the $d_{xy}$ orbital (which contributes mostly to the $\gamma$ band).
Since there is no reason to expect a degeneracy between even and odd-parity components for $d_{xy}$ (because it is not quasi-1D), we assume that only one component, the even one, is present on that orbital.
To sum up, the proposed scenario is the following: an even-parity $\Delta_e$ component appears at the first transition on all three orbitals, and an odd-parity component $\Delta_o$ appears at a second transition only on the quasi-1D orbitals.

Before discussing in more details the form $\Delta_e$ and $\Delta_o$ could take within a three-orbital model, we can already discuss the general properties of a state of the type $\Delta_e + i \Delta_o$.
Such a state has several desirable features as a candidate for multi-component superconductivity in \SRO{}.
First, the accidental degeneracy between the two components has a microscopic justification based on the small parameter $t_y/t_x$. 
Second, the OP is still nodal even though it forms a complex linear combination, since both components have ``cosine nodes'' at $k_y = \pm \pi/2$ (resp. $k_x = \pm \pi/2$) for $d_{zx}$ (resp. for $d_{zy}$). (The presence of nodes in the superconducting gap is well established\cite{NishiZakiEA00,PhysRevLett.85.4775,LupienEA01,HassingerEA17,Sharma5222}, although their location remains controversial.)
Third, the fact that $|\Delta_e(\bk)|^2 = |\Delta_o(\bk)|^2$ everywhere on the Fermi surface leads to a parametrically small second specific heat jump, as required by recent measurements~\cite{2019arXiv190607597L}.

Another problem facing most proposals of time-reversal symmetry-breaking order parameters is that it contradicts the absence of measurable edge currents revealed by magnetometry measurements\cite{PhysRevB.81.214501}.
Even though several effects have been predicted to reduce these currents \cite{PhysRevB.91.094507,PhysRevB.90.224519,PhysRevB.90.134521,ScaffidiSimon15}, this remains a challenge for most OPs with TRS breaking, like $p+i p$ or $d+id$.
By contrast, a state of the type $\Delta_e + i \Delta_o$ provides a natural way of breaking time-reversal symmetry without having edge currents (in a centrosymmetric crystal).
Indeed, the gradient terms which usually lead to spontaneous edge currents are not allowed in this case since they do not respect parity\footnote{These gradient terms become important close to sample edges, domain walls, or defects, in the vicinity of which the order parameter is not spatially homogeneous.}:
\bea
F \not\supset \int d\bx \ (\partial_x \psi_e^*) (\partial_y \psi_o) + \text{c.c.}
\eea
where $\psi_o$ and $\psi_e$ are the components as defined in Eq.~\ref{ComplexComb}.

If no edge currents are expected, what is the manifestation of time-reversal symmetry breaking for mixed parity states? It actually manifests itself through the spin degree of freedom, rather than the orbital one. Indeed, mixed even-odd parity superconductors experience a spontaneous magnetization at any non-homogenities, like domain walls, edges, and defects \cite{doi:10.7566/JPSJ.83.044712,2017arXiv171105241Y,Robins_2018}. The intuition is that the relative $i$ phase is between two different spin (or rather helicity) components, rather than two different orbital components (e.g. $p_x$ and $p_y$). The orientation of the spontaneous magnetization depends on the orientation of $\vec{d}$ and of the inhomogeneity.
For example, for a state of the type $d_{x^2-y^2} + i (p_x - p_y) \hat{z}$ (as proposed below), the following term would be allowed by symmetry \cite{2017arXiv171105241Y}:
\bea
F \supset \int d\bx \ m_z \psi_e^* (\partial_x + \partial_y) \psi_o 
\eea
where $m_z$ is the $z$ component of the magnetization.
This term would create a spontaneous $\hat{z}$ magnetization localized around inhomogeneities of the order parameter.

More generally, a magnetization $\vec{m}$ localized around extended defects like domain walls or dislocations could explain the presence of a signal in muSR \cite{LukeEA98, 2020arXiv200108152G} (regardless of the orientation of $\vec{m}$) and in the Kerr effect \cite{XiaEA06} (as long as $\vec{m}$ has an out-of-plane component).
It could also explain the absence of a signal in scanning SQUID magnetometry measurements \cite{PhysRevB.81.214501}, since a surface magnetization does not produce stray fields.
Note also that the scale of the magnetization would depend on microscopic details and is probably directly related to the strength of spin-orbit coupling.
An additional phenomenon to consider when studying muSR is that the muon itself could create a local magnetization in a mixed-parity superconductor, since it can be seen as a charged defect.

Besides, the behavior of superconductivity in \SRO{} under $[1,0,0]$ strain could also be explained by the current scenario.
First, no cusp of $T_c$ at zero strain is expected for an accidental degeneracy \cite{HicksEA14}.
Second, it is natural to expect the even-parity component to undergo a large increase of $T_c$ as the $\gamma$ band approaches the van Hove singularity, since the even-parity component is by assumption non-zero on that band, and is anti-nodal at the van Hove point \cite{SteppkeEA17}.
By contrast, one would only expect a small variation of the onset temperature for the odd-parity component since it only resides on the quasi-1D bands, which are comparatively little affected by strain. This would be consistent with the small variation of the onset temperature of the muSR signal observed in Ref.~\cite{2020arXiv200108152G}.

Further, the presence of an odd-parity, pseudo-spin triplet component would help explain a number of experiments which have been interpreted that way, like Josephson junction tunneling\cite{Nelson1151,Kidwingira1267,PhysRevB.95.224509}, the observation of half-quantum vortices \cite{Jang186}, and \SRO-ferromagnet heterostructures \cite{PhysRevB.100.024516}.

In the next two subsections, we will discuss in more details the different ways in which the two components $\Delta_e$ and $\Delta_o$ obtained in the simple model of Section II could be incorporated into a three-orbital model of \SRO{}.
We will also examine the implications for other experiments, namely the measurement of the Knight shift\cite{PustogowEA19,IshidaEA19}, and of the jump in elastic moduli\cite{lupien2002ultrasound,2020arXiv200206130G,2020arXiv200205916B}.
\subsection{Nature of the even-parity component}
Assuming that a gap of the form $\cos(k_y)$ (resp. $\cos(k_x)$) is favored on $d_{zx}$ (resp. $d_{zy}$), there remains the question of the relative phase between the gaps in the two orbitals.
If this phase is $+1$ (resp. ($-1$)), the resulting gap is in the $A_{1g}$ (resp. $B_{1g}$) representation:
\bea
A_{1g} &: (\Delta_{e,d_{zx}},\Delta_{e,d_{zy}})=(\cos(k_y),\cos(k_x)) \\
B_{1g} &: (\Delta_{e,d_{zx}},\Delta_{e,d_{zy}})=(\cos(k_y),-\cos(k_x))
\eea
where $\Delta_{e,d_{zx}}$ (resp. $\Delta_{e,d_{zy}}$) is the even-parity component on the $d_{zx}$ (resp. $d_{zy}$) orbital.
The difference between $A_{1g}$ ($s'$) and $B_{1g}$ ($d_{x^2-y^2}$) only becomes important along the diagonals ($[1,1,0]$ and $[1,-1,0]$ directions), since the $B_{1g}$ gap has symmetry-imposed nodes along the diagonals, while the $A_{1g}$ gap does not.
By contrast, the ``cosine'' nodes at $k_x = \pm \pi/2$ and $k_y = \pm \pi/2$ are present for both $A_{1g}$ and $B_{1g}$.

Within a two-orbital model, the splitting between $A_{1g}$ and $B_{1g}$ is a ``second order effect'', since it only depends on the hybridization between the two orbitals, which is mostly localized in a small region along the diagonals.
In fact, a close competition between these states has been reported in previous work, even in three-orbital models \cite{RomerEA19,PhysRevB.101.064507}.
Both $s'$ and $d_{x^2-y^2}$ should therefore be considered as candidates for the even-parity component.

\subsection{Nature of the odd-parity component}
We expect the odd-parity order to only arise on the $d_{zx}$ and $d_{zy}$ orbitals, since the degeneracy between odd- and even-parity states relies on the quasi-1D limit.
Starting from the $\mathrm{sign}(k_x) \cos(k_y)$ form found in the single orbital model, two choices have to be made: the spin orientation of Cooper pairs (parametrized by $\vec{d}$) on each orbital, and the relative phase of the OPs between the two orbitals.
Each choice corresponds to a different $D_{4h}$ representation:
\begin{widetext}
\bea
E_u : (\vec{d}_{d_{zx}},\vec{d}_{d_{zy}}) &= \hat{z} (\eta_x \ \mathrm{sign}(k_x) \cos(k_y), \eta_y \ \mathrm{sign}(k_y) \cos(k_x)) \\
A_{1u} : (\vec{d}_{d_{zx}},\vec{d}_{d_{zy}}) &= (\hat{x} \ \mathrm{sign}(k_x) \cos(k_y), \hat{y} \ \mathrm{sign}(k_y) \cos(k_x) ) \\
A_{2u} : (\vec{d}_{d_{zx}},\vec{d}_{d_{zy}}) &= (\hat{y} \ \mathrm{sign}(k_x) \cos(k_y), -\hat{x} \ \mathrm{sign}(k_y) \cos(k_x) ) \\
B_{1u} : (\vec{d}_{d_{zx}},\vec{d}_{d_{zy}}) &= (\hat{x} \ \mathrm{sign}(k_x) \cos(k_y), -\hat{y} \ \mathrm{sign}(k_y) \cos(k_x) ) \\
B_{2u} : (\vec{d}_{d_{zx}},\vec{d}_{d_{zy}}) &= (\hat{y} \ \mathrm{sign}(k_x) \cos(k_y), \hat{x} \ \mathrm{sign}(k_y) \cos(k_x) )
\eea
\end{widetext}
where $\vec{d}_{d_{zx}}$ (resp. $\vec{d}_{d_{zy}}$) is the $\vec{d}$ vector on the $d_{zx}$ (resp. $d_{zy}$ orbital), and where $\eta_x$ and $\eta_y$ are free parameters.
All these representations are degenerate for the $SU(2)$-symmetric single-orbital toy model considered in Section II.
They would however be split by spin-orbit coupling in a realistic model, as studied in previous work (see Ref.~\cite{PhysRevB.101.064507} and references therein). 
We will take here a phenomenological approach and discuss the different representations at the light of available experimental results.

\subsubsection{$E_u$ state}
The favored $E_u$ state can either be $[1,0,0]$-nematic $\{p_x,p_y\}$, $[1,1,0]$-nematic $p_x \pm p_y$, or chiral $p_x \pm i p_y$.
Whereas a chiral state is usually favored since it does not have any symmetry-imposed nodes, the situation is different here due to the presence of the even-parity component.
It is indeed favorable for both the $p_x$ and the $p_y$ components to have a relative $\pm i$ phase with respect to the even-parity component (in order to form a unitary state), which is of course incompatible with having a relative $i$ phase between $p_x$ and $p_y$. 
A nematic state could therefore be favored due to the presence of the even-parity component.
Since a $100$-nematic state seems unlikely due to the fact that it would only gap out one of the two quasi-1D orbitals, the most likely scenario would be a $[1,1,0]$-nematic state: $p_x \pm p_y$.
Combining this with the above candidates for the even-component, the OP would be of the form $d_{x^2-y^2} + i (p_x \pm p_y) \hat{z}$ or $s' + i (p_x \pm p_y) \hat{z}$.

Neglecting spin-orbit coupling and assuming an equal amplitude of singlet and triplet components on the $\alpha$ and $\beta$ bands at $T=0$, we can obtain an estimate of the residual spin susceptibilities based on Section III.B:
\bea
\frac{\chi_{\parallel}(T=0)}{\chi_N} &= \frac12 \frac{\rho_{\alpha,\beta}}{\rho} \simeq 0.2\\
\frac{\chi_{\perp}(T=0)}{\chi_N} &= 0
\label{KnightShiftChiral}
\eea
for in-plane and out-of-plane magnetic fields, respectively, and where $\rho_{\alpha,\beta}$ is the density of states (DOS) at the Fermi level for the alpha and beta bands, and $\rho=\rho_{\alpha,\beta}+\rho_{\gamma}$ is the total DOS. Quantum oscillation measurements give $\frac{\rho_{\alpha,\beta}}{\rho} \simeq 0.4$ ~\cite{MackenzieMaeno03}.
A residual susceptibility of $0.2$ was consistent with earlier Knight shift measurements\cite{PustogowEA19,IshidaEA19}, but is inconsistent with the upper bound of $0.1$ recently reported by Chronister et al.
Based on our current estimate for the residual susceptibility, a mixed $\Delta_e + i E_u$ state is therefore inconsistent with the latest Knight shift experiments.
A more accurate estimate of $\chi/\chi_N$ based on a microscopic calculation with spin-orbit coupling and multi-band effects is however warranted before the possibility of such a state is discarded altogether.

Regarding ultrasound experiments, an $E_u$ component would explain the presence of a jump in the $B_{2g}$ elastic modulus \cite{lupien2002ultrasound,2020arXiv200206130G,2020arXiv200205916B}, but could also potentially have a jump in the $B_{1g}$ channel, which was not observed (although there could be some microscopic reasons why the $B_{1g}$ jump has a smaller prefactor).

\subsubsection{Helical states ($A_{1u},A_{2u},B_{1u},B_{2u}$)}
Helical states have a $\vec{d}$ vector that rotates in plane as one moves around the Fermi surface.
An accurate calculation of the spin susceptibility is beyond the scope of this work, but we can already obtain an estimate as follows.
Assuming an approximately isotropic orientation of $\vec{d}$ within the plane, helical states would have the following residual spin susceptibilities:
\bea
\frac{\chi_{\parallel}(T=0)}{\chi_N} &= \frac14 \frac{\rho_{\alpha,\beta}}{\rho} \simeq 0.1\\
\frac{\chi_{\perp}(T=0)}{\chi_N} &= \frac12 \frac{\rho_{\alpha,\beta}}{\rho} \simeq 0.2
\label{KnightShiftHelical}
\eea
for in-plane and out-of-plane magnetic fields, respectively.
To the best of our knowledge, these values are compatible with current NMR experiments, but could potentially be disproved by further measurements~\cite{PustogowEA19,IshidaEA19}.

It does not seem possible at this point to explain a jump in the $B_{2g}$ elastic modulus without invoking an accidental combination of two different helical states, like $B_{1u}$ and $A_{2u}$. However, a thorough analysis of possible couplings between elasticity and mixed even-odd order parameters might reveal other possibilities, especially if inhomogeneities of the order parameter are taken into account.

A necessary (though not sufficient \cite{PhysRevLett.108.157001}) criterion to see a Kerr signal is to break time-reversal symmetry and all vertical mirror planes \cite{Kapitulnik_2009}.
If inhomogeneities (e.g. domain walls) can be invoked to break certain mirror symmetries, the Kerr signal cannot discriminate between different helical states.
However, if one requires all vertical mirror symmetries to be broken by the bulk order parameter, the presence of a Kerr signal imposes restrictions on the possible helical states: assuming that the even component is in $A_{1g}$ or $B_{1g}$, only combinations of the type $A_{1g} + i  A_{1u}$ or $B_{1g} + i B_{1u}$ would break all vertical mirrors.

\section{Discussion}
We have established an accidental degeneracy between even-parity ($\Delta_e = \cos(k_y)$) and odd-parity ($\Delta_o = \cos(k_y) \mathrm{sign}(k_x)$) superconducting orders in the quasi-1D limit ($t_y/t_x \rightarrow 0$) of the Hubbard model, in the weak $U$ limit.
Moving away from the purely 1D limit creates a small splitting between these orders by favoring the even-parity one.
A Ginzburg-Landau analysis then revealed that a linear combination of the type $\Delta_e + i \Delta_o$ can become favorable at a second transition.
Remarkably, the degenerate orders have essentially the same gap magnitude over the entire Fermi surface, leading to a parametrically small $b'$ coefficient in the Ginzburg-Landau free energy.
This leads to a parametrically small specific heat jump at the second transition.

In Section IV, we assumed that this mechanism is at play on the quasi-1D orbitals of \SRO{}, and we analyzed the consequences for experiments.
  A state of the type $\Delta_e + i \Delta_o$ has several desirable features. 
It explains the presence of nodes~\cite{NishiZakiEA00,PhysRevLett.85.4775,LupienEA01,HassingerEA17,Sharma5222} in a time-reversal symmetry breaking state, and it predicts a parametrically small specific heat jump \cite{2019arXiv190607597L}.
  It also reconciles the breaking of time-reversal symmetry~\cite{LukeEA98, XiaEA06} with the absence of edge currents~\cite{PhysRevB.81.214501}.
  Further, the presence of an odd-parity, pseudo-spin triplet component would help explain a number of measurements which have been interpreted as such\cite{Nelson1151,Kidwingira1267,Jang186,PhysRevB.100.024516}.
  
  Whereas our solution of the single orbital Hubbard model is exact, its application to \SRO{} was purely empirical, since we do not have a microscopic justification for neglecting inter-orbital effects.
  These effects have been studied extensively in the literature\cite{Damascelli08,Raghu2EA10,Wang_2013,HuoEA13,DamascelliEA14,ScaffidiEA14,PhysRevB.101.064507,PhysRevResearch.1.033108}, and can often impact crucially the predictions of theoretical models.
Our ambition with this work was much smaller: we wanted to find a toy model which exhibits a second transition to a TRS breaking state with a parametrically small specific heat jump, which we have found.
 A more realistic calculation which includes multiple orbitals and spin-orbit coupling would be necessary to go beyond this proof of principle.
 The main effect which could create substantial splitting between even and odd-parity SC orders is inter-orbital interaction, as already observed in Ref.\cite{Raghu2EA10}. A thorough study of the fate of this degeneracy as a function of $J/U$ is therefore warranted (where $J$ is the Hund's coupling and $U$ is the intra-orbital Hubbard interaction).

 Besides, the quasi-1D regime of the square lattice Hubbard model is relevant to a variety of materials, including   Bechgaard salts~\cite{bourbonnais2008physics,PhysRevB.80.214531,WeejeeEA14} and Li$_{0.9}$Mo$_6$O$_{17}$~\cite{WeejeeEA15}.
 This model can be generalized to the case of longer range interaction and finite $U/t$, which leads to a variety of interesting superconducting phases, including odd-frequency superconductivity and Fulde-Ferrell-Larkin-Ovchinnikov phases \cite{PhysRevB.83.140509,PhysRevLett.102.016403,PhysRevB.70.060502,PhysRevB.79.174507}.
  Moving beyond the quasi-1D regime, an accidental degeneracy between even and odd-parity superconductivity is an interesting possibility to consider \footnote{This possibility was actually first mentioned by Leggett in 1975 \cite{RevModPhys.47.331}, but no microscopic model exhibiting this behavior was known at the time. We thank Catherine Kallin for bringing this point to our attention.}, in the context of \SRO{} and of other systems. 
In fact, the proximity to a quantum critical point was shown to provide another mechanism for a nearly degenerate pairing in even and odd channels \cite{PhysRevLett.114.097001,PhysRevLett.115.207002,PhysRevLett.117.217003,PhysRevB.92.125108,PhysRevB.93.134512,PhysRevLett.118.227001,PhysRevB.99.144507}.

One defining feature of a mixed-parity state is of course the breaking of inversion symmetry, which could be probed by non-linear optical effects like second-harmonic generation \cite{Zhao:2016aa,Zhao:2017aa,PhysRevB.100.220501}.
Another way to measure a breaking of inversion symmetry is provided by phase-sensitive measurements which probe opposite sides of the sample \cite{Nelson1151}.
A study of the nature of edge modes in a mixed-parity state could also reveal interesting properties, and could be compared with existing experimental data~\cite{KashiwayaEA11}.
Finally, the most direct way to put the present proposal to the test is probably the Knight shift \cite{PustogowEA19,IshidaEA19}: The presence of a spin-triplet component could be disproved if a residual susceptibility smaller than the ones predicted in Eq.~\ref{KnightShiftChiral} or \ref{KnightShiftHelical} was measured.

 As we were completing this work, we received a manuscript
by Chronister \textit{et al.}~\cite{Chronister2020} reporting new Knight shift measurements in \SRO{}. These measurements provide a more constraining upper bound on the spin susceptibility of the condensate than previous work. Based on our estimates for the residual susceptibility, the results of Chronister et al do not rule out the possibility of a mixed-parity order parameter.

\acknowledgements{
We would like to acknowledge helpful discussions with Stuart Brown, Felix Flicker, Clifford Hicks, Wen Huang, Catherine Kallin, Andrew Mackenzie, Srinivas Raghu, Henrik Roising, Joerg Schmalian, and Steven Simon.
We acknowledge the support of the Natural Sciences and Engineering Research Council of Canada (NSERC), in particular the Discovery Grant [RGPIN-2020-05842], the Accelerator Supplements [RGPAS-2020-00060], and the Discovery Launch Supplement [DGECR-2020-00222].
 }

\bibliography{Quasi_1D_paper}

\begin{thebibliography}{106}%
\makeatletter
\providecommand \@ifxundefined [1]{%
 \@ifx{#1\undefined}
}%
\providecommand \@ifnum [1]{%
 \ifnum #1\expandafter \@firstoftwo
 \else \expandafter \@secondoftwo
 \fi
}%
\providecommand \@ifx [1]{%
 \ifx #1\expandafter \@firstoftwo
 \else \expandafter \@secondoftwo
 \fi
}%
\providecommand \natexlab [1]{#1}%
\providecommand \enquote  [1]{``#1''}%
\providecommand \bibnamefont  [1]{#1}%
\providecommand \bibfnamefont [1]{#1}%
\providecommand \citenamefont [1]{#1}%
\providecommand \href@noop [0]{\@secondoftwo}%
\providecommand \href [0]{\begingroup \@sanitize@url \@href}%
\providecommand \@href[1]{\@@startlink{#1}\@@href}%
\providecommand \@@href[1]{\endgroup#1\@@endlink}%
\providecommand \@sanitize@url [0]{\catcode `\\12\catcode `\$12\catcode
  `\&12\catcode `\#12\catcode `\^12\catcode `\_12\catcode `\%12\relax}%
\providecommand \@@startlink[1]{}%
\providecommand \@@endlink[0]{}%
\providecommand \url  [0]{\begingroup\@sanitize@url \@url }%
\providecommand \@url [1]{\endgroup\@href {#1}{\urlprefix }}%
\providecommand \urlprefix  [0]{URL }%
\providecommand \Eprint [0]{\href }%
\providecommand \doibase [0]{http://dx.doi.org/}%
\providecommand \selectlanguage [0]{\@gobble}%
\providecommand \bibinfo  [0]{\@secondoftwo}%
\providecommand \bibfield  [0]{\@secondoftwo}%
\providecommand \translation [1]{[#1]}%
\providecommand \BibitemOpen [0]{}%
\providecommand \bibitemStop [0]{}%
\providecommand \bibitemNoStop [0]{.\EOS\space}%
\providecommand \EOS [0]{\spacefactor3000\relax}%
\providecommand \BibitemShut  [1]{\csname bibitem#1\endcsname}%
\let\auto@bib@innerbib\@empty
\bibitem [{\citenamefont {Leggett}(1975)}]{RevModPhys.47.331}%
  \BibitemOpen
  \bibfield  {author} {\bibinfo {author} {\bibfnamefont {A.~J.}\ \bibnamefont
  {Leggett}},\ }\href {\doibase 10.1103/RevModPhys.47.331} {\bibfield
  {journal} {\bibinfo  {journal} {Rev. Mod. Phys.}\ }\textbf {\bibinfo {volume}
  {47}},\ \bibinfo {pages} {331} (\bibinfo {year} {1975})}\BibitemShut
  {NoStop}%
\bibitem [{\citenamefont {Volovik}(2003)}]{volovik2003universe}%
  \BibitemOpen
  \bibfield  {author} {\bibinfo {author} {\bibfnamefont {G.~E.}\ \bibnamefont
  {Volovik}},\ }\href@noop {} {\emph {\bibinfo {title} {The universe in a
  helium droplet}}},\ Vol.\ \bibinfo {volume} {117}\ (\bibinfo  {publisher}
  {Oxford University Press on Demand},\ \bibinfo {year} {2003})\BibitemShut
  {NoStop}%
\bibitem [{\citenamefont {Qi}\ and\ \citenamefont
  {Zhang}(2011)}]{RevModPhys.83.1057}%
  \BibitemOpen
  \bibfield  {author} {\bibinfo {author} {\bibfnamefont {X.-L.}\ \bibnamefont
  {Qi}}\ and\ \bibinfo {author} {\bibfnamefont {S.-C.}\ \bibnamefont {Zhang}},\
  }\href {\doibase 10.1103/RevModPhys.83.1057} {\bibfield  {journal} {\bibinfo
  {journal} {Rev. Mod. Phys.}\ }\textbf {\bibinfo {volume} {83}},\ \bibinfo
  {pages} {1057} (\bibinfo {year} {2011})}\BibitemShut {NoStop}%
\bibitem [{\citenamefont {Maiti}\ and\ \citenamefont
  {Chubukov}(2013)}]{PhysRevB.87.144511}%
  \BibitemOpen
  \bibfield  {author} {\bibinfo {author} {\bibfnamefont {S.}~\bibnamefont
  {Maiti}}\ and\ \bibinfo {author} {\bibfnamefont {A.~V.}\ \bibnamefont
  {Chubukov}},\ }\href {\doibase 10.1103/PhysRevB.87.144511} {\bibfield
  {journal} {\bibinfo  {journal} {Phys. Rev. B}\ }\textbf {\bibinfo {volume}
  {87}},\ \bibinfo {pages} {144511} (\bibinfo {year} {2013})}\BibitemShut
  {NoStop}%
\bibitem [{\citenamefont {Weng}\ \emph {et~al.}(2016)\citenamefont {Weng},
  \citenamefont {Zhang}, \citenamefont {Smidman}, \citenamefont {Shang},
  \citenamefont {Quintanilla}, \citenamefont {Annett}, \citenamefont {Nicklas},
  \citenamefont {Pang}, \citenamefont {Jiao}, \citenamefont {Jiang},
  \citenamefont {Chen}, \citenamefont {Steglich},\ and\ \citenamefont
  {Yuan}}]{PhysRevLett.117.027001}%
  \BibitemOpen
  \bibfield  {author} {\bibinfo {author} {\bibfnamefont {Z.~F.}\ \bibnamefont
  {Weng}}, \bibinfo {author} {\bibfnamefont {J.~L.}\ \bibnamefont {Zhang}},
  \bibinfo {author} {\bibfnamefont {M.}~\bibnamefont {Smidman}}, \bibinfo
  {author} {\bibfnamefont {T.}~\bibnamefont {Shang}}, \bibinfo {author}
  {\bibfnamefont {J.}~\bibnamefont {Quintanilla}}, \bibinfo {author}
  {\bibfnamefont {J.~F.}\ \bibnamefont {Annett}}, \bibinfo {author}
  {\bibfnamefont {M.}~\bibnamefont {Nicklas}}, \bibinfo {author} {\bibfnamefont
  {G.~M.}\ \bibnamefont {Pang}}, \bibinfo {author} {\bibfnamefont
  {L.}~\bibnamefont {Jiao}}, \bibinfo {author} {\bibfnamefont {W.~B.}\
  \bibnamefont {Jiang}}, \bibinfo {author} {\bibfnamefont {Y.}~\bibnamefont
  {Chen}}, \bibinfo {author} {\bibfnamefont {F.}~\bibnamefont {Steglich}}, \
  and\ \bibinfo {author} {\bibfnamefont {H.~Q.}\ \bibnamefont {Yuan}},\ }\href
  {\doibase 10.1103/PhysRevLett.117.027001} {\bibfield  {journal} {\bibinfo
  {journal} {Phys. Rev. Lett.}\ }\textbf {\bibinfo {volume} {117}},\ \bibinfo
  {pages} {027001} (\bibinfo {year} {2016})}\BibitemShut {NoStop}%
\bibitem [{\citenamefont {{Kivelson}}\ \emph {et~al.}(2020)\citenamefont
  {{Kivelson}}, \citenamefont {{Yuan}}, \citenamefont {{Ramshaw}},\ and\
  \citenamefont {{Thomale}}}]{2020arXiv200200016K}%
  \BibitemOpen
  \bibfield  {author} {\bibinfo {author} {\bibfnamefont {S.~A.}\ \bibnamefont
  {{Kivelson}}}, \bibinfo {author} {\bibfnamefont {A.~C.}\ \bibnamefont
  {{Yuan}}}, \bibinfo {author} {\bibfnamefont {B.~J.}\ \bibnamefont
  {{Ramshaw}}}, \ and\ \bibinfo {author} {\bibfnamefont {R.}~\bibnamefont
  {{Thomale}}},\ }\href@noop {} {\bibfield  {journal} {\bibinfo  {journal}
  {arXiv e-prints}\ ,\ \bibinfo {eid} {arXiv:2002.00016}} (\bibinfo {year}
  {2020})},\ \Eprint {http://arxiv.org/abs/2002.00016} {arXiv:2002.00016
  [cond-mat.supr-con]} \BibitemShut {NoStop}%
\bibitem [{\citenamefont {Maeno}\ \emph {et~al.}(1994)\citenamefont {Maeno},
  \citenamefont {Hashimoto}, \citenamefont {Yoshida}, \citenamefont
  {Nishizaki}, \citenamefont {Fujita}, \citenamefont {Bednorz},\ and\
  \citenamefont {Lichtenberg}}]{MaenoEA94}%
  \BibitemOpen
  \bibfield  {author} {\bibinfo {author} {\bibfnamefont {Y.}~\bibnamefont
  {Maeno}}, \bibinfo {author} {\bibfnamefont {H.}~\bibnamefont {Hashimoto}},
  \bibinfo {author} {\bibfnamefont {K.}~\bibnamefont {Yoshida}}, \bibinfo
  {author} {\bibfnamefont {S.}~\bibnamefont {Nishizaki}}, \bibinfo {author}
  {\bibfnamefont {T.}~\bibnamefont {Fujita}}, \bibinfo {author} {\bibfnamefont
  {J.~G.}\ \bibnamefont {Bednorz}}, \ and\ \bibinfo {author} {\bibfnamefont
  {F.}~\bibnamefont {Lichtenberg}},\ }\href {\doibase 10.1038/372532a0}
  {\bibfield  {journal} {\bibinfo  {journal} {Nature}\ }\textbf {\bibinfo
  {volume} {372}},\ \bibinfo {pages} {532 } (\bibinfo {year}
  {1994})}\BibitemShut {NoStop}%
\bibitem [{\citenamefont {Rice}\ and\ \citenamefont
  {Sigrist}(1995)}]{RiceSigrist95}%
  \BibitemOpen
  \bibfield  {author} {\bibinfo {author} {\bibfnamefont {T.~M.}\ \bibnamefont
  {Rice}}\ and\ \bibinfo {author} {\bibfnamefont {M.}~\bibnamefont {Sigrist}},\
  }\href {\doibase 10.1088/0953-8984/7/47/002} {\bibfield  {journal} {\bibinfo
  {journal} {J. Phys. Condens. Matter}\ }\textbf {\bibinfo {volume} {7}},\
  \bibinfo {pages} {L643} (\bibinfo {year} {1995})}\BibitemShut {NoStop}%
\bibitem [{\citenamefont {Baskaran}(1996)}]{BASKARAN1996490}%
  \BibitemOpen
  \bibfield  {author} {\bibinfo {author} {\bibfnamefont {G.}~\bibnamefont
  {Baskaran}},\ }\href {\doibase https://doi.org/10.1016/0921-4526(96)00155-X}
  {\bibfield  {journal} {\bibinfo  {journal} {Physica B: Condensed Matter}\
  }\textbf {\bibinfo {volume} {223-224}},\ \bibinfo {pages} {490 } (\bibinfo
  {year} {1996})},\ \bibinfo {note} {proceedings of the International
  Conference on Strongly Correlated Electron Systems}\BibitemShut {NoStop}%
\bibitem [{\citenamefont {Mackenzie}\ and\ \citenamefont
  {Maeno}(2003)}]{MackenzieMaeno03}%
  \BibitemOpen
  \bibfield  {author} {\bibinfo {author} {\bibfnamefont {A.~P.}\ \bibnamefont
  {Mackenzie}}\ and\ \bibinfo {author} {\bibfnamefont {Y.}~\bibnamefont
  {Maeno}},\ }\href {\doibase 10.1103/RevModPhys.75.657} {\bibfield  {journal}
  {\bibinfo  {journal} {Rev. Mod. Phys.}\ }\textbf {\bibinfo {volume} {75}},\
  \bibinfo {pages} {657} (\bibinfo {year} {2003})}\BibitemShut {NoStop}%
\bibitem [{\citenamefont {Maeno}\ \emph {et~al.}(2012)\citenamefont {Maeno},
  \citenamefont {Kittaka}, \citenamefont {Nomura}, \citenamefont {Yonezawa},\
  and\ \citenamefont {Ishida}}]{MaenoEA12}%
  \BibitemOpen
  \bibfield  {author} {\bibinfo {author} {\bibfnamefont {Y.}~\bibnamefont
  {Maeno}}, \bibinfo {author} {\bibfnamefont {S.}~\bibnamefont {Kittaka}},
  \bibinfo {author} {\bibfnamefont {T.}~\bibnamefont {Nomura}}, \bibinfo
  {author} {\bibfnamefont {S.}~\bibnamefont {Yonezawa}}, \ and\ \bibinfo
  {author} {\bibfnamefont {K.}~\bibnamefont {Ishida}},\ }\href {\doibase
  10.1143/JPSJ.81.011009} {\bibfield  {journal} {\bibinfo  {journal} {J. Phys.
  Soc. Jpn.}\ }\textbf {\bibinfo {volume} {81}},\ \bibinfo {pages} {011009}
  (\bibinfo {year} {2012})}\BibitemShut {NoStop}%
\bibitem [{\citenamefont {Kallin}\ and\ \citenamefont
  {Berlinsky}(2009)}]{Kallin_2009}%
  \BibitemOpen
  \bibfield  {author} {\bibinfo {author} {\bibfnamefont {C.}~\bibnamefont
  {Kallin}}\ and\ \bibinfo {author} {\bibfnamefont {A.~J.}\ \bibnamefont
  {Berlinsky}},\ }\href {\doibase 10.1088/0953-8984/21/16/164210} {\bibfield
  {journal} {\bibinfo  {journal} {Journal of Physics: Condensed Matter}\
  }\textbf {\bibinfo {volume} {21}},\ \bibinfo {pages} {164210} (\bibinfo
  {year} {2009})}\BibitemShut {NoStop}%
\bibitem [{\citenamefont {Kallin}(2012)}]{Kallin12}%
  \BibitemOpen
  \bibfield  {author} {\bibinfo {author} {\bibfnamefont {C.}~\bibnamefont
  {Kallin}},\ }\href {\doibase 10.1088/0034-4885/75/4/042501} {\bibfield
  {journal} {\bibinfo  {journal} {Rep. Prog. Phys.}\ }\textbf {\bibinfo
  {volume} {75}},\ \bibinfo {pages} {042501} (\bibinfo {year}
  {2012})}\BibitemShut {NoStop}%
\bibitem [{\citenamefont {Mackenzie}\ \emph {et~al.}(2017)\citenamefont
  {Mackenzie}, \citenamefont {Scaffidi}, \citenamefont {Hicks},\ and\
  \citenamefont {Maeno}}]{MackenzieEA17}%
  \BibitemOpen
  \bibfield  {author} {\bibinfo {author} {\bibfnamefont {A.~P.}\ \bibnamefont
  {Mackenzie}}, \bibinfo {author} {\bibfnamefont {T.}~\bibnamefont {Scaffidi}},
  \bibinfo {author} {\bibfnamefont {C.~W.}\ \bibnamefont {Hicks}}, \ and\
  \bibinfo {author} {\bibfnamefont {Y.}~\bibnamefont {Maeno}},\ }\href
  {\doibase 10.1038/s41535-017-0045-4} {\bibfield  {journal} {\bibinfo
  {journal} {npj Quantum Materials}\ }\textbf {\bibinfo {volume} {2}} (\bibinfo
  {year} {2017}),\ 10.1038/s41535-017-0045-4}\BibitemShut {NoStop}%
\bibitem [{\citenamefont {Damascelli}\ \emph {et~al.}(2000)\citenamefont
  {Damascelli}, \citenamefont {Lu}, \citenamefont {Shen}, \citenamefont
  {Armitage}, \citenamefont {Ronning}, \citenamefont {Feng}, \citenamefont
  {Kim}, \citenamefont {Shen}, \citenamefont {Kimura}, \citenamefont {Tokura},
  \citenamefont {Mao},\ and\ \citenamefont {Maeno}}]{DamascelliEA00}%
  \BibitemOpen
  \bibfield  {author} {\bibinfo {author} {\bibfnamefont {A.}~\bibnamefont
  {Damascelli}}, \bibinfo {author} {\bibfnamefont {D.~H.}\ \bibnamefont {Lu}},
  \bibinfo {author} {\bibfnamefont {K.~M.}\ \bibnamefont {Shen}}, \bibinfo
  {author} {\bibfnamefont {N.~P.}\ \bibnamefont {Armitage}}, \bibinfo {author}
  {\bibfnamefont {F.}~\bibnamefont {Ronning}}, \bibinfo {author} {\bibfnamefont
  {D.~L.}\ \bibnamefont {Feng}}, \bibinfo {author} {\bibfnamefont
  {C.}~\bibnamefont {Kim}}, \bibinfo {author} {\bibfnamefont {Z.-X.}\
  \bibnamefont {Shen}}, \bibinfo {author} {\bibfnamefont {T.}~\bibnamefont
  {Kimura}}, \bibinfo {author} {\bibfnamefont {Y.}~\bibnamefont {Tokura}},
  \bibinfo {author} {\bibfnamefont {Z.~Q.}\ \bibnamefont {Mao}}, \ and\
  \bibinfo {author} {\bibfnamefont {Y.}~\bibnamefont {Maeno}},\ }\href
  {\doibase 10.1103/PhysRevLett.85.5194} {\bibfield  {journal} {\bibinfo
  {journal} {Phys. Rev. Lett.}\ }\textbf {\bibinfo {volume} {85}},\ \bibinfo
  {pages} {5194} (\bibinfo {year} {2000})}\BibitemShut {NoStop}%
\bibitem [{\citenamefont {Bergemann}\ \emph {et~al.}(2000)\citenamefont
  {Bergemann}, \citenamefont {Julian}, \citenamefont {Mackenzie}, \citenamefont
  {NishiZaki},\ and\ \citenamefont {Maeno}}]{BergmannEA00}%
  \BibitemOpen
  \bibfield  {author} {\bibinfo {author} {\bibfnamefont {C.}~\bibnamefont
  {Bergemann}}, \bibinfo {author} {\bibfnamefont {S.~R.}\ \bibnamefont
  {Julian}}, \bibinfo {author} {\bibfnamefont {A.~P.}\ \bibnamefont
  {Mackenzie}}, \bibinfo {author} {\bibfnamefont {S.}~\bibnamefont
  {NishiZaki}}, \ and\ \bibinfo {author} {\bibfnamefont {Y.}~\bibnamefont
  {Maeno}},\ }\href {\doibase 10.1103/PhysRevLett.84.2662} {\bibfield
  {journal} {\bibinfo  {journal} {Phys. Rev. Lett.}\ }\textbf {\bibinfo
  {volume} {84}},\ \bibinfo {pages} {2662} (\bibinfo {year}
  {2000})}\BibitemShut {NoStop}%
\bibitem [{\citenamefont {Bergemann}\ \emph {et~al.}(2003)\citenamefont
  {Bergemann}, \citenamefont {Mackenzie}, \citenamefont {Julian}, \citenamefont
  {Forsythe},\ and\ \citenamefont {Ohmichi}}]{BergmannEA03}%
  \BibitemOpen
  \bibfield  {author} {\bibinfo {author} {\bibfnamefont {C.}~\bibnamefont
  {Bergemann}}, \bibinfo {author} {\bibfnamefont {A.~P.}\ \bibnamefont
  {Mackenzie}}, \bibinfo {author} {\bibfnamefont {S.~R.}\ \bibnamefont
  {Julian}}, \bibinfo {author} {\bibfnamefont {D.}~\bibnamefont {Forsythe}}, \
  and\ \bibinfo {author} {\bibfnamefont {E.}~\bibnamefont {Ohmichi}},\ }\href
  {\doibase 10.1080/00018730310001621737} {\bibfield  {journal} {\bibinfo
  {journal} {Adv. Phys.}\ }\textbf {\bibinfo {volume} {52}},\ \bibinfo {pages}
  {639} (\bibinfo {year} {2003})}\BibitemShut {NoStop}%
\bibitem [{\citenamefont {Tamai}\ \emph {et~al.}(2019)\citenamefont {Tamai},
  \citenamefont {Zingl}, \citenamefont {Rozbicki}, \citenamefont {Cappelli},
  \citenamefont {Ricc\`o}, \citenamefont {de~la Torre}, \citenamefont
  {McKeown~Walker}, \citenamefont {Bruno}, \citenamefont {King}, \citenamefont
  {Meevasana}, \citenamefont {Shi}, \citenamefont
  {Radovi\ifmmode~\acute{c}\else \'{c}\fi{}}, \citenamefont {Plumb},
  \citenamefont {Gibbs}, \citenamefont {Mackenzie}, \citenamefont {Berthod},
  \citenamefont {Strand}, \citenamefont {Kim}, \citenamefont {Georges},\ and\
  \citenamefont {Baumberger}}]{TamaiEA18}%
  \BibitemOpen
  \bibfield  {author} {\bibinfo {author} {\bibfnamefont {A.}~\bibnamefont
  {Tamai}}, \bibinfo {author} {\bibfnamefont {M.}~\bibnamefont {Zingl}},
  \bibinfo {author} {\bibfnamefont {E.}~\bibnamefont {Rozbicki}}, \bibinfo
  {author} {\bibfnamefont {E.}~\bibnamefont {Cappelli}}, \bibinfo {author}
  {\bibfnamefont {S.}~\bibnamefont {Ricc\`o}}, \bibinfo {author} {\bibfnamefont
  {A.}~\bibnamefont {de~la Torre}}, \bibinfo {author} {\bibfnamefont
  {S.}~\bibnamefont {McKeown~Walker}}, \bibinfo {author} {\bibfnamefont
  {F.~Y.}\ \bibnamefont {Bruno}}, \bibinfo {author} {\bibfnamefont {P.~D.~C.}\
  \bibnamefont {King}}, \bibinfo {author} {\bibfnamefont {W.}~\bibnamefont
  {Meevasana}}, \bibinfo {author} {\bibfnamefont {M.}~\bibnamefont {Shi}},
  \bibinfo {author} {\bibfnamefont {M.}~\bibnamefont
  {Radovi\ifmmode~\acute{c}\else \'{c}\fi{}}}, \bibinfo {author} {\bibfnamefont
  {N.~C.}\ \bibnamefont {Plumb}}, \bibinfo {author} {\bibfnamefont {A.~S.}\
  \bibnamefont {Gibbs}}, \bibinfo {author} {\bibfnamefont {A.~P.}\ \bibnamefont
  {Mackenzie}}, \bibinfo {author} {\bibfnamefont {C.}~\bibnamefont {Berthod}},
  \bibinfo {author} {\bibfnamefont {H.~U.~R.}\ \bibnamefont {Strand}}, \bibinfo
  {author} {\bibfnamefont {M.}~\bibnamefont {Kim}}, \bibinfo {author}
  {\bibfnamefont {A.}~\bibnamefont {Georges}}, \ and\ \bibinfo {author}
  {\bibfnamefont {F.}~\bibnamefont {Baumberger}},\ }\href {\doibase
  10.1103/PhysRevX.9.021048} {\bibfield  {journal} {\bibinfo  {journal} {Phys.
  Rev. X}\ }\textbf {\bibinfo {volume} {9}},\ \bibinfo {pages} {021048}
  (\bibinfo {year} {2019})}\BibitemShut {NoStop}%
\bibitem [{\citenamefont {Luke}\ \emph {et~al.}(1998)\citenamefont {Luke},
  \citenamefont {Fudamoto}, \citenamefont {Kojima}, \citenamefont {Larkin},
  \citenamefont {Merrin}, \citenamefont {Nachumi}, \citenamefont {Uemura},
  \citenamefont {Maeno}, \citenamefont {Mao}, \citenamefont {Mori},
  \citenamefont {Nakamura},\ and\ \citenamefont {Sigrist}}]{LukeEA98}%
  \BibitemOpen
  \bibfield  {author} {\bibinfo {author} {\bibfnamefont {G.~M.}\ \bibnamefont
  {Luke}}, \bibinfo {author} {\bibfnamefont {Y.}~\bibnamefont {Fudamoto}},
  \bibinfo {author} {\bibfnamefont {K.~M.}\ \bibnamefont {Kojima}}, \bibinfo
  {author} {\bibfnamefont {M.~I.}\ \bibnamefont {Larkin}}, \bibinfo {author}
  {\bibfnamefont {J.}~\bibnamefont {Merrin}}, \bibinfo {author} {\bibfnamefont
  {B.}~\bibnamefont {Nachumi}}, \bibinfo {author} {\bibfnamefont {Y.~J.}\
  \bibnamefont {Uemura}}, \bibinfo {author} {\bibfnamefont {Y.}~\bibnamefont
  {Maeno}}, \bibinfo {author} {\bibfnamefont {Z.~Q.}\ \bibnamefont {Mao}},
  \bibinfo {author} {\bibfnamefont {Y.}~\bibnamefont {Mori}}, \bibinfo {author}
  {\bibfnamefont {H.}~\bibnamefont {Nakamura}}, \ and\ \bibinfo {author}
  {\bibfnamefont {M.}~\bibnamefont {Sigrist}},\ }\href {\doibase 10.1038/29038}
  {\bibfield  {journal} {\bibinfo  {journal} {Nature}\ }\textbf {\bibinfo
  {volume} {394}},\ \bibinfo {pages} {0028} (\bibinfo {year}
  {1998})}\BibitemShut {NoStop}%
\bibitem [{\citenamefont {Xia}\ \emph {et~al.}(2006)\citenamefont {Xia},
  \citenamefont {Maeno}, \citenamefont {Beyersdorf}, \citenamefont {Fejer},\
  and\ \citenamefont {Kapitulnik}}]{XiaEA06}%
  \BibitemOpen
  \bibfield  {author} {\bibinfo {author} {\bibfnamefont {J.}~\bibnamefont
  {Xia}}, \bibinfo {author} {\bibfnamefont {Y.}~\bibnamefont {Maeno}}, \bibinfo
  {author} {\bibfnamefont {P.~T.}\ \bibnamefont {Beyersdorf}}, \bibinfo
  {author} {\bibfnamefont {M.~M.}\ \bibnamefont {Fejer}}, \ and\ \bibinfo
  {author} {\bibfnamefont {A.}~\bibnamefont {Kapitulnik}},\ }\href {\doibase
  10.1103/PhysRevLett.97.167002} {\bibfield  {journal} {\bibinfo  {journal}
  {Phys. Rev. Lett.}\ }\textbf {\bibinfo {volume} {97}},\ \bibinfo {pages}
  {167002} (\bibinfo {year} {2006})}\BibitemShut {NoStop}%
\bibitem [{\citenamefont {{Grinenko}}\ \emph {et~al.}(2020)\citenamefont
  {{Grinenko}}, \citenamefont {{Ghosh}}, \citenamefont {{Sarkar}},
  \citenamefont {{Orain}}, \citenamefont {{Nikitin}}, \citenamefont
  {{Elender}}, \citenamefont {{Das}}, \citenamefont {{Guguchia}}, \citenamefont
  {{Br{\"u}ckner}}, \citenamefont {{Barber}}, \citenamefont {{Park}},
  \citenamefont {{Kikugawa}}, \citenamefont {{Sokolov}}, \citenamefont
  {{Bobowski}}, \citenamefont {{Miyoshi}}, \citenamefont {{Maeno}},
  \citenamefont {{Mackenzie}}, \citenamefont {{Luetkens}}, \citenamefont
  {{Hicks}},\ and\ \citenamefont {{Klauss}}}]{2020arXiv200108152G}%
  \BibitemOpen
  \bibfield  {author} {\bibinfo {author} {\bibfnamefont {V.}~\bibnamefont
  {{Grinenko}}}, \bibinfo {author} {\bibfnamefont {S.}~\bibnamefont {{Ghosh}}},
  \bibinfo {author} {\bibfnamefont {R.}~\bibnamefont {{Sarkar}}}, \bibinfo
  {author} {\bibfnamefont {J.-C.}\ \bibnamefont {{Orain}}}, \bibinfo {author}
  {\bibfnamefont {A.}~\bibnamefont {{Nikitin}}}, \bibinfo {author}
  {\bibfnamefont {M.}~\bibnamefont {{Elender}}}, \bibinfo {author}
  {\bibfnamefont {D.}~\bibnamefont {{Das}}}, \bibinfo {author} {\bibfnamefont
  {Z.}~\bibnamefont {{Guguchia}}}, \bibinfo {author} {\bibfnamefont
  {F.}~\bibnamefont {{Br{\"u}ckner}}}, \bibinfo {author} {\bibfnamefont
  {M.~E.}\ \bibnamefont {{Barber}}}, \bibinfo {author} {\bibfnamefont
  {J.}~\bibnamefont {{Park}}}, \bibinfo {author} {\bibfnamefont
  {N.}~\bibnamefont {{Kikugawa}}}, \bibinfo {author} {\bibfnamefont {D.~A.}\
  \bibnamefont {{Sokolov}}}, \bibinfo {author} {\bibfnamefont {J.~S.}\
  \bibnamefont {{Bobowski}}}, \bibinfo {author} {\bibfnamefont
  {T.}~\bibnamefont {{Miyoshi}}}, \bibinfo {author} {\bibfnamefont
  {Y.}~\bibnamefont {{Maeno}}}, \bibinfo {author} {\bibfnamefont {A.~P.}\
  \bibnamefont {{Mackenzie}}}, \bibinfo {author} {\bibfnamefont
  {H.}~\bibnamefont {{Luetkens}}}, \bibinfo {author} {\bibfnamefont {C.~W.}\
  \bibnamefont {{Hicks}}}, \ and\ \bibinfo {author} {\bibfnamefont {H.-H.}\
  \bibnamefont {{Klauss}}},\ }\href@noop {} {\bibfield  {journal} {\bibinfo
  {journal} {arXiv e-prints}\ ,\ \bibinfo {eid} {arXiv:2001.08152}} (\bibinfo
  {year} {2020})},\ \Eprint {http://arxiv.org/abs/2001.08152} {arXiv:2001.08152
  [cond-mat.supr-con]} \BibitemShut {NoStop}%
\bibitem [{\citenamefont {Lupien}(2002)}]{lupien2002ultrasound}%
  \BibitemOpen
  \bibfield  {author} {\bibinfo {author} {\bibfnamefont {C.}~\bibnamefont
  {Lupien}},\ }\href@noop {} {\emph {\bibinfo {title} {Ultrasound attenuation
  in the unconventional superconductor Sr2RuO4}}}\ (\bibinfo  {publisher} {PhD
  thesis},\ \bibinfo {year} {2002})\BibitemShut {NoStop}%
\bibitem [{\citenamefont {{Ghosh}}\ \emph {et~al.}(2020)\citenamefont
  {{Ghosh}}, \citenamefont {{Shekhter}}, \citenamefont {{Jerzembeck}},
  \citenamefont {{Kikugawa}}, \citenamefont {{Sokolov}}, \citenamefont
  {{Brando}}, \citenamefont {{Mackenzie}}, \citenamefont {{Hicks}},\ and\
  \citenamefont {{Ramshaw}}}]{2020arXiv200206130G}%
  \BibitemOpen
  \bibfield  {author} {\bibinfo {author} {\bibfnamefont {S.}~\bibnamefont
  {{Ghosh}}}, \bibinfo {author} {\bibfnamefont {A.}~\bibnamefont {{Shekhter}}},
  \bibinfo {author} {\bibfnamefont {F.}~\bibnamefont {{Jerzembeck}}}, \bibinfo
  {author} {\bibfnamefont {N.}~\bibnamefont {{Kikugawa}}}, \bibinfo {author}
  {\bibfnamefont {D.~A.}\ \bibnamefont {{Sokolov}}}, \bibinfo {author}
  {\bibfnamefont {M.}~\bibnamefont {{Brando}}}, \bibinfo {author}
  {\bibfnamefont {A.~P.}\ \bibnamefont {{Mackenzie}}}, \bibinfo {author}
  {\bibfnamefont {C.~W.}\ \bibnamefont {{Hicks}}}, \ and\ \bibinfo {author}
  {\bibfnamefont {B.~J.}\ \bibnamefont {{Ramshaw}}},\ }\href@noop {} {\bibfield
   {journal} {\bibinfo  {journal} {arXiv e-prints}\ ,\ \bibinfo {eid}
  {arXiv:2002.06130}} (\bibinfo {year} {2020})},\ \Eprint
  {http://arxiv.org/abs/2002.06130} {arXiv:2002.06130 [cond-mat.supr-con]}
  \BibitemShut {NoStop}%
\bibitem [{\citenamefont {{Benhabib}}\ \emph {et~al.}(2020)\citenamefont
  {{Benhabib}}, \citenamefont {{Lupien}}, \citenamefont {{Paul}}, \citenamefont
  {{Berges}}, \citenamefont {{Dion}}, \citenamefont {{Nardone}}, \citenamefont
  {{Zitouni}}, \citenamefont {{Mao}}, \citenamefont {{Maeno}}, \citenamefont
  {{Georges}}, \citenamefont {{Taillefer}},\ and\ \citenamefont
  {{Proust}}}]{2020arXiv200205916B}%
  \BibitemOpen
  \bibfield  {author} {\bibinfo {author} {\bibfnamefont {S.}~\bibnamefont
  {{Benhabib}}}, \bibinfo {author} {\bibfnamefont {C.}~\bibnamefont
  {{Lupien}}}, \bibinfo {author} {\bibfnamefont {I.}~\bibnamefont {{Paul}}},
  \bibinfo {author} {\bibfnamefont {L.}~\bibnamefont {{Berges}}}, \bibinfo
  {author} {\bibfnamefont {M.}~\bibnamefont {{Dion}}}, \bibinfo {author}
  {\bibfnamefont {M.}~\bibnamefont {{Nardone}}}, \bibinfo {author}
  {\bibfnamefont {A.}~\bibnamefont {{Zitouni}}}, \bibinfo {author}
  {\bibfnamefont {Z.~Q.}\ \bibnamefont {{Mao}}}, \bibinfo {author}
  {\bibfnamefont {Y.}~\bibnamefont {{Maeno}}}, \bibinfo {author} {\bibfnamefont
  {A.}~\bibnamefont {{Georges}}}, \bibinfo {author} {\bibfnamefont
  {L.}~\bibnamefont {{Taillefer}}}, \ and\ \bibinfo {author} {\bibfnamefont
  {C.}~\bibnamefont {{Proust}}},\ }\href@noop {} {\bibfield  {journal}
  {\bibinfo  {journal} {arXiv e-prints}\ ,\ \bibinfo {eid} {arXiv:2002.05916}}
  (\bibinfo {year} {2020})},\ \Eprint {http://arxiv.org/abs/2002.05916}
  {arXiv:2002.05916 [cond-mat.supr-con]} \BibitemShut {NoStop}%
\bibitem [{\citenamefont {Pustogow}\ \emph {et~al.}(2019)\citenamefont
  {Pustogow}, \citenamefont {Luo}, \citenamefont {Chronister}, \citenamefont
  {Su}, \citenamefont {Sokolov}, \citenamefont {Jerzembeck}, \citenamefont
  {Mackenzie}, \citenamefont {Hicks}, \citenamefont {Kikugawa}, \citenamefont
  {Raghu},\ and\ \citenamefont {Bauer}}]{PustogowEA19}%
  \BibitemOpen
  \bibfield  {author} {\bibinfo {author} {\bibfnamefont {A.}~\bibnamefont
  {Pustogow}}, \bibinfo {author} {\bibfnamefont {Y.}~\bibnamefont {Luo}},
  \bibinfo {author} {\bibfnamefont {A.}~\bibnamefont {Chronister}}, \bibinfo
  {author} {\bibfnamefont {Y.-S.}\ \bibnamefont {Su}}, \bibinfo {author}
  {\bibfnamefont {D.~A.}\ \bibnamefont {Sokolov}}, \bibinfo {author}
  {\bibfnamefont {F.}~\bibnamefont {Jerzembeck}}, \bibinfo {author}
  {\bibfnamefont {A.~P.}\ \bibnamefont {Mackenzie}}, \bibinfo {author}
  {\bibfnamefont {C.~W.}\ \bibnamefont {Hicks}}, \bibinfo {author}
  {\bibfnamefont {N.}~\bibnamefont {Kikugawa}}, \bibinfo {author}
  {\bibfnamefont {S.}~\bibnamefont {Raghu}}, \ and\ \bibinfo {author}
  {\bibfnamefont {S.~E.}\ \bibnamefont {Bauer}, \bibfnamefont {E.~D.~Brown}},\
  }\href {\doibase 10.1038/s41586-019-1596-2} {\bibfield  {journal} {\bibinfo
  {journal} {Nature}\ }\textbf {\bibinfo {volume} {574}},\ \bibinfo {pages}
  {72} (\bibinfo {year} {2019})}\BibitemShut {NoStop}%
\bibitem [{\citenamefont {Ishida}\ \emph {et~al.}(2019)\citenamefont {Ishida},
  \citenamefont {Manago},\ and\ \citenamefont {Maeno}}]{IshidaEA19}%
  \BibitemOpen
  \bibfield  {author} {\bibinfo {author} {\bibfnamefont {K.}~\bibnamefont
  {Ishida}}, \bibinfo {author} {\bibfnamefont {M.}~\bibnamefont {Manago}}, \
  and\ \bibinfo {author} {\bibfnamefont {Y.}~\bibnamefont {Maeno}},\ }\href
  {https://arxiv.org/abs/1907.12236} {\bibfield  {journal} {\bibinfo  {journal}
  {arXiv:1907.12236 [cond-mat.supr-con]}\ } (\bibinfo {year}
  {2019})}\BibitemShut {NoStop}%
\bibitem [{\citenamefont {Huang}\ and\ \citenamefont
  {Yao}(2018)}]{PhysRevLett.121.157002}%
  \BibitemOpen
  \bibfield  {author} {\bibinfo {author} {\bibfnamefont {W.}~\bibnamefont
  {Huang}}\ and\ \bibinfo {author} {\bibfnamefont {H.}~\bibnamefont {Yao}},\
  }\href {\doibase 10.1103/PhysRevLett.121.157002} {\bibfield  {journal}
  {\bibinfo  {journal} {Phys. Rev. Lett.}\ }\textbf {\bibinfo {volume} {121}},\
  \bibinfo {pages} {157002} (\bibinfo {year} {2018})}\BibitemShut {NoStop}%
\bibitem [{\citenamefont {Ramires}\ and\ \citenamefont
  {Sigrist}(2019)}]{RamiresSigrist19}%
  \BibitemOpen
  \bibfield  {author} {\bibinfo {author} {\bibfnamefont {A.}~\bibnamefont
  {Ramires}}\ and\ \bibinfo {author} {\bibfnamefont {M.}~\bibnamefont
  {Sigrist}},\ }\href {\doibase 10.1103/PhysRevB.100.104501} {\bibfield
  {journal} {\bibinfo  {journal} {Phys. Rev. B}\ }\textbf {\bibinfo {volume}
  {100}},\ \bibinfo {pages} {104501} (\bibinfo {year} {2019})}\BibitemShut
  {NoStop}%
\bibitem [{\citenamefont {R{\o}mer}\ \emph {et~al.}(2019)\citenamefont
  {R{\o}mer}, \citenamefont {Scherer}, \citenamefont {Eremin}, \citenamefont
  {Hirschfeld},\ and\ \citenamefont {Andersen}}]{RomerEA19}%
  \BibitemOpen
  \bibfield  {author} {\bibinfo {author} {\bibfnamefont {A.~T.}\ \bibnamefont
  {R{\o}mer}}, \bibinfo {author} {\bibfnamefont {D.~D.}\ \bibnamefont
  {Scherer}}, \bibinfo {author} {\bibfnamefont {I.~M.}\ \bibnamefont {Eremin}},
  \bibinfo {author} {\bibfnamefont {P.~J.}\ \bibnamefont {Hirschfeld}}, \ and\
  \bibinfo {author} {\bibfnamefont {B.~M.}\ \bibnamefont {Andersen}},\ }\href
  {https://arxiv.org/abs/1905.04782} {\bibfield  {journal} {\bibinfo  {journal}
  {arXiv:1905.047822 [cond-mat.supr-con]}\ } (\bibinfo {year}
  {2019})}\BibitemShut {NoStop}%
\bibitem [{\citenamefont {{Gyeol Suh}}\ \emph {et~al.}(2019)\citenamefont
  {{Gyeol Suh}}, \citenamefont {{Menke}}, \citenamefont {{Brydon}},
  \citenamefont {{Timm}}, \citenamefont {{Ramires}},\ and\ \citenamefont
  {{Agterberg}}}]{2019arXiv191209525G}%
  \BibitemOpen
  \bibfield  {author} {\bibinfo {author} {\bibfnamefont {H.}~\bibnamefont
  {{Gyeol Suh}}}, \bibinfo {author} {\bibfnamefont {H.}~\bibnamefont
  {{Menke}}}, \bibinfo {author} {\bibfnamefont {P.~M.~R.}\ \bibnamefont
  {{Brydon}}}, \bibinfo {author} {\bibfnamefont {C.}~\bibnamefont {{Timm}}},
  \bibinfo {author} {\bibfnamefont {A.}~\bibnamefont {{Ramires}}}, \ and\
  \bibinfo {author} {\bibfnamefont {D.~F.}\ \bibnamefont {{Agterberg}}},\
  }\href@noop {} {\bibfield  {journal} {\bibinfo  {journal} {arXiv e-prints}\
  ,\ \bibinfo {eid} {arXiv:1912.09525}} (\bibinfo {year} {2019})},\ \Eprint
  {http://arxiv.org/abs/1912.09525} {arXiv:1912.09525 [cond-mat.supr-con]}
  \BibitemShut {NoStop}%
\bibitem [{\citenamefont {Huang}\ \emph
  {et~al.}(2019{\natexlab{a}})\citenamefont {Huang}, \citenamefont {Zhou},\
  and\ \citenamefont {Yao}}]{HuangEA19}%
  \BibitemOpen
  \bibfield  {author} {\bibinfo {author} {\bibfnamefont {W.}~\bibnamefont
  {Huang}}, \bibinfo {author} {\bibfnamefont {Y.}~\bibnamefont {Zhou}}, \ and\
  \bibinfo {author} {\bibfnamefont {H.}~\bibnamefont {Yao}},\ }\href
  {https://arxiv.org/abs/1905.03523} {\bibfield  {journal} {\bibinfo  {journal}
  {arXiv:1905.03523 [cond-mat.supr-con]}\ } (\bibinfo {year}
  {2019}{\natexlab{a}})}\BibitemShut {NoStop}%
\bibitem [{\citenamefont {Huang}\ \emph
  {et~al.}(2019{\natexlab{b}})\citenamefont {Huang}, \citenamefont {Zhou},\
  and\ \citenamefont {Yao}}]{HuangNematicEA19}%
  \BibitemOpen
  \bibfield  {author} {\bibinfo {author} {\bibfnamefont {W.}~\bibnamefont
  {Huang}}, \bibinfo {author} {\bibfnamefont {Y.}~\bibnamefont {Zhou}}, \ and\
  \bibinfo {author} {\bibfnamefont {H.}~\bibnamefont {Yao}},\ }\href
  {https://arxiv.org/abs/1901.07041} {\bibfield  {journal} {\bibinfo  {journal}
  {arXiv:1901.07041 [cond-mat.supr-con]}\ } (\bibinfo {year}
  {2019}{\natexlab{b}})}\BibitemShut {NoStop}%
\bibitem [{\citenamefont {{Li}}\ \emph {et~al.}(2019)\citenamefont {{Li}},
  \citenamefont {{Kikugawa}}, \citenamefont {{Sokolov}}, \citenamefont
  {{Jerzembeck}}, \citenamefont {{Gibbs}}, \citenamefont {{Maeno}},
  \citenamefont {{Hicks}}, \citenamefont {{Nicklas}},\ and\ \citenamefont
  {{Mackenzie}}}]{2019arXiv190607597L}%
  \BibitemOpen
  \bibfield  {author} {\bibinfo {author} {\bibfnamefont {Y.~S.}\ \bibnamefont
  {{Li}}}, \bibinfo {author} {\bibfnamefont {N.}~\bibnamefont {{Kikugawa}}},
  \bibinfo {author} {\bibfnamefont {D.~A.}\ \bibnamefont {{Sokolov}}}, \bibinfo
  {author} {\bibfnamefont {F.}~\bibnamefont {{Jerzembeck}}}, \bibinfo {author}
  {\bibfnamefont {A.~S.}\ \bibnamefont {{Gibbs}}}, \bibinfo {author}
  {\bibfnamefont {Y.}~\bibnamefont {{Maeno}}}, \bibinfo {author} {\bibfnamefont
  {C.~W.}\ \bibnamefont {{Hicks}}}, \bibinfo {author} {\bibfnamefont
  {M.}~\bibnamefont {{Nicklas}}}, \ and\ \bibinfo {author} {\bibfnamefont
  {A.~P.}\ \bibnamefont {{Mackenzie}}},\ }\href@noop {} {\bibfield  {journal}
  {\bibinfo  {journal} {arXiv e-prints}\ ,\ \bibinfo {eid} {arXiv:1906.07597}}
  (\bibinfo {year} {2019})},\ \Eprint {http://arxiv.org/abs/1906.07597}
  {arXiv:1906.07597 [cond-mat.supr-con]} \BibitemShut {NoStop}%
\bibitem [{\citenamefont {Raghu}\ \emph
  {et~al.}(2010{\natexlab{a}})\citenamefont {Raghu}, \citenamefont
  {Kapitulnik},\ and\ \citenamefont {Kivelson}}]{Raghu2EA10}%
  \BibitemOpen
  \bibfield  {author} {\bibinfo {author} {\bibfnamefont {S.}~\bibnamefont
  {Raghu}}, \bibinfo {author} {\bibfnamefont {A.}~\bibnamefont {Kapitulnik}}, \
  and\ \bibinfo {author} {\bibfnamefont {S.~A.}\ \bibnamefont {Kivelson}},\
  }\href {\doibase 10.1103/PhysRevLett.105.136401} {\bibfield  {journal}
  {\bibinfo  {journal} {Phys. Rev. Lett.}\ }\textbf {\bibinfo {volume} {105}},\
  \bibinfo {pages} {136401} (\bibinfo {year} {2010}{\natexlab{a}})}\BibitemShut
  {NoStop}%
\bibitem [{\citenamefont {Kohn}\ and\ \citenamefont
  {Luttinger}(1965)}]{KohnLuttinger65}%
  \BibitemOpen
  \bibfield  {author} {\bibinfo {author} {\bibfnamefont {W.}~\bibnamefont
  {Kohn}}\ and\ \bibinfo {author} {\bibfnamefont {J.~M.}\ \bibnamefont
  {Luttinger}},\ }\href {\doibase 10.1103/PhysRevLett.15.524} {\bibfield
  {journal} {\bibinfo  {journal} {Phys. Rev. Lett.}\ }\textbf {\bibinfo
  {volume} {15}},\ \bibinfo {pages} {524} (\bibinfo {year} {1965})}\BibitemShut
  {NoStop}%
\bibitem [{\citenamefont {Baranov}\ and\ \citenamefont
  {Kagan}(1992)}]{BaranovEA92}%
  \BibitemOpen
  \bibfield  {author} {\bibinfo {author} {\bibfnamefont {M.~A.}\ \bibnamefont
  {Baranov}}\ and\ \bibinfo {author} {\bibfnamefont {M.~Y.}\ \bibnamefont
  {Kagan}},\ }\href {\doibase 10.1007/BF01313830} {\bibfield  {journal}
  {\bibinfo  {journal} {Z. Phys. B}\ }\textbf {\bibinfo {volume} {86}},\
  \bibinfo {pages} {237} (\bibinfo {year} {1992})}\BibitemShut {NoStop}%
\bibitem [{\citenamefont {Kagan}\ and\ \citenamefont
  {Chubukov}(1989)}]{KaganChubukov89}%
  \BibitemOpen
  \bibfield  {author} {\bibinfo {author} {\bibfnamefont {M.~Y.}\ \bibnamefont
  {Kagan}}\ and\ \bibinfo {author} {\bibfnamefont {A.}~\bibnamefont
  {Chubukov}},\ }\href
  {http://www.jetpletters.ac.ru/ps/1134/article_17165.shtml} {\bibfield
  {journal} {\bibinfo  {journal} {JETP Lett.}\ }\textbf {\bibinfo {volume}
  {50}},\ \bibinfo {pages} {517} (\bibinfo {year} {1989})}\BibitemShut
  {NoStop}%
\bibitem [{\citenamefont {Chubukov}\ and\ \citenamefont
  {Lu}(1992)}]{ChubukovEA92}%
  \BibitemOpen
  \bibfield  {author} {\bibinfo {author} {\bibfnamefont {A.~V.}\ \bibnamefont
  {Chubukov}}\ and\ \bibinfo {author} {\bibfnamefont {J.~P.}\ \bibnamefont
  {Lu}},\ }\href {\doibase 10.1103/PhysRevB.46.11163} {\bibfield  {journal}
  {\bibinfo  {journal} {Phys. Rev. B}\ }\textbf {\bibinfo {volume} {46}},\
  \bibinfo {pages} {11163} (\bibinfo {year} {1992})}\BibitemShut {NoStop}%
\bibitem [{\citenamefont {Baranov}\ \emph {et~al.}(1992)\citenamefont
  {Baranov}, \citenamefont {Chubukov},\ and\ \citenamefont
  {Yu.~Kagan}}]{BaranovChubukovEA92}%
  \BibitemOpen
  \bibfield  {author} {\bibinfo {author} {\bibfnamefont {M.~A.}\ \bibnamefont
  {Baranov}}, \bibinfo {author} {\bibfnamefont {A.~V.}\ \bibnamefont
  {Chubukov}}, \ and\ \bibinfo {author} {\bibfnamefont {M.}~\bibnamefont
  {Yu.~Kagan}},\ }\href {\doibase 10.1142/S0217979292001249} {\bibfield
  {journal} {\bibinfo  {journal} {Int. J. Mod. Phys. A}\ }\textbf {\bibinfo
  {volume} {06}},\ \bibinfo {pages} {2471} (\bibinfo {year}
  {1992})}\BibitemShut {NoStop}%
\bibitem [{\citenamefont {Chubukov}(1993)}]{ChubukovEA93}%
  \BibitemOpen
  \bibfield  {author} {\bibinfo {author} {\bibfnamefont {A.~V.}\ \bibnamefont
  {Chubukov}},\ }\href {\doibase 10.1103/PhysRevB.48.1097} {\bibfield
  {journal} {\bibinfo  {journal} {Phys. Rev. B}\ }\textbf {\bibinfo {volume}
  {48}},\ \bibinfo {pages} {1097} (\bibinfo {year} {1993})}\BibitemShut
  {NoStop}%
\bibitem [{\citenamefont {Fukazawa}\ and\ \citenamefont
  {Yamada}(2002)}]{HironoEA02}%
  \BibitemOpen
  \bibfield  {author} {\bibinfo {author} {\bibfnamefont {H.}~\bibnamefont
  {Fukazawa}}\ and\ \bibinfo {author} {\bibfnamefont {K.}~\bibnamefont
  {Yamada}},\ }\href {\doibase 10.1143/JPSJ.71.1541} {\bibfield  {journal}
  {\bibinfo  {journal} {J. Phys. Soc. Jpn.}\ }\textbf {\bibinfo {volume}
  {71}},\ \bibinfo {pages} {1541} (\bibinfo {year} {2002})}\BibitemShut
  {NoStop}%
\bibitem [{\citenamefont {Hlubina}(1999)}]{Hlubina99}%
  \BibitemOpen
  \bibfield  {author} {\bibinfo {author} {\bibfnamefont {R.}~\bibnamefont
  {Hlubina}},\ }\href {\doibase 10.1103/PhysRevB.59.9600} {\bibfield  {journal}
  {\bibinfo  {journal} {Phys. Rev. B}\ }\textbf {\bibinfo {volume} {59}},\
  \bibinfo {pages} {9600} (\bibinfo {year} {1999})}\BibitemShut {NoStop}%
\bibitem [{\citenamefont {Raghu}\ \emph
  {et~al.}(2010{\natexlab{b}})\citenamefont {Raghu}, \citenamefont {Kivelson},\
  and\ \citenamefont {Scalapino}}]{RaghuEA10}%
  \BibitemOpen
  \bibfield  {author} {\bibinfo {author} {\bibfnamefont {S.}~\bibnamefont
  {Raghu}}, \bibinfo {author} {\bibfnamefont {S.~A.}\ \bibnamefont {Kivelson}},
  \ and\ \bibinfo {author} {\bibfnamefont {D.~J.}\ \bibnamefont {Scalapino}},\
  }\href {\doibase 10.1103/PhysRevB.81.224505} {\bibfield  {journal} {\bibinfo
  {journal} {Phys. Rev. B}\ }\textbf {\bibinfo {volume} {81}},\ \bibinfo
  {pages} {224505} (\bibinfo {year} {2010}{\natexlab{b}})}\BibitemShut
  {NoStop}%
\bibitem [{\citenamefont {Cho}\ \emph {et~al.}(2013)\citenamefont {Cho},
  \citenamefont {Thomale}, \citenamefont {Raghu},\ and\ \citenamefont
  {Kivelson}}]{WeejeeEA14}%
  \BibitemOpen
  \bibfield  {author} {\bibinfo {author} {\bibfnamefont {W.}~\bibnamefont
  {Cho}}, \bibinfo {author} {\bibfnamefont {R.}~\bibnamefont {Thomale}},
  \bibinfo {author} {\bibfnamefont {S.}~\bibnamefont {Raghu}}, \ and\ \bibinfo
  {author} {\bibfnamefont {S.~A.}\ \bibnamefont {Kivelson}},\ }\href {\doibase
  10.1103/PhysRevB.88.064505} {\bibfield  {journal} {\bibinfo  {journal} {Phys.
  Rev. B}\ }\textbf {\bibinfo {volume} {88}},\ \bibinfo {pages} {064505}
  (\bibinfo {year} {2013})}\BibitemShut {NoStop}%
\bibitem [{\citenamefont {Scaffidi}\ \emph {et~al.}(2014)\citenamefont
  {Scaffidi}, \citenamefont {Romers},\ and\ \citenamefont
  {Simon}}]{ScaffidiEA14}%
  \BibitemOpen
  \bibfield  {author} {\bibinfo {author} {\bibfnamefont {T.}~\bibnamefont
  {Scaffidi}}, \bibinfo {author} {\bibfnamefont {J.~C.}\ \bibnamefont
  {Romers}}, \ and\ \bibinfo {author} {\bibfnamefont {S.~H.}\ \bibnamefont
  {Simon}},\ }\href {\doibase 10.1103/PhysRevB.89.220510} {\bibfield  {journal}
  {\bibinfo  {journal} {Phys. Rev. B}\ }\textbf {\bibinfo {volume} {89}},\
  \bibinfo {pages} {220510} (\bibinfo {year} {2014})}\BibitemShut {NoStop}%
\bibitem [{\citenamefont {\ifmmode~\check{S}\else \v{S}\fi{}imkovic}\ \emph
  {et~al.}(2016)\citenamefont {\ifmmode~\check{S}\else \v{S}\fi{}imkovic},
  \citenamefont {Liu}, \citenamefont {Deng},\ and\ \citenamefont
  {Kozik}}]{SimkovicEA16}%
  \BibitemOpen
  \bibfield  {author} {\bibinfo {author} {\bibfnamefont {F.}~\bibnamefont
  {\ifmmode~\check{S}\else \v{S}\fi{}imkovic}}, \bibinfo {author}
  {\bibfnamefont {X.-W.}\ \bibnamefont {Liu}}, \bibinfo {author} {\bibfnamefont
  {Y.}~\bibnamefont {Deng}}, \ and\ \bibinfo {author} {\bibfnamefont
  {E.}~\bibnamefont {Kozik}},\ }\href {\doibase 10.1103/PhysRevB.94.085106}
  {\bibfield  {journal} {\bibinfo  {journal} {Phys. Rev. B}\ }\textbf {\bibinfo
  {volume} {94}},\ \bibinfo {pages} {085106} (\bibinfo {year}
  {2016})}\BibitemShut {NoStop}%
\bibitem [{\citenamefont {Scaffidi}(2017)}]{Scaffidi2017}%
  \BibitemOpen
  \bibfield  {author} {\bibinfo {author} {\bibfnamefont {T.}~\bibnamefont
  {Scaffidi}},\ }\href {https://books.google.com/books?id=bQpKswEACAAJ} {\emph
  {\bibinfo {title} {{Weak-Coupling Theory of Topological Superconductivity:
  The Case of Strontium Ruthenate}}}},\ Springer Theses\ (\bibinfo  {publisher}
  {Springer International Publishing},\ \bibinfo {year} {2017})\BibitemShut
  {NoStop}%
\bibitem [{\citenamefont {R\o{}ising}\ \emph {et~al.}(2018)\citenamefont
  {R\o{}ising}, \citenamefont {Flicker}, \citenamefont {Scaffidi},\ and\
  \citenamefont {Simon}}]{RoisingEA18}%
  \BibitemOpen
  \bibfield  {author} {\bibinfo {author} {\bibfnamefont {H.~S.}\ \bibnamefont
  {R\o{}ising}}, \bibinfo {author} {\bibfnamefont {F.}~\bibnamefont {Flicker}},
  \bibinfo {author} {\bibfnamefont {T.}~\bibnamefont {Scaffidi}}, \ and\
  \bibinfo {author} {\bibfnamefont {S.~H.}\ \bibnamefont {Simon}},\ }\href
  {\doibase 10.1103/PhysRevB.98.224515} {\bibfield  {journal} {\bibinfo
  {journal} {Phys. Rev. B}\ }\textbf {\bibinfo {volume} {98}},\ \bibinfo
  {pages} {224515} (\bibinfo {year} {2018})}\BibitemShut {NoStop}%
\bibitem [{\citenamefont {Giamarchi}\ and\ \citenamefont
  {Press}(2004)}]{giamarchi2004quantum}%
  \BibitemOpen
  \bibfield  {author} {\bibinfo {author} {\bibfnamefont {T.}~\bibnamefont
  {Giamarchi}}\ and\ \bibinfo {author} {\bibfnamefont {O.~U.}\ \bibnamefont
  {Press}},\ }\href {https://books.google.ca/books?id=1MwTDAAAQBAJ} {\emph
  {\bibinfo {title} {Quantum Physics in One Dimension}}},\ International Series
  of Monogr\ (\bibinfo  {publisher} {Clarendon Press},\ \bibinfo {year}
  {2004})\BibitemShut {NoStop}%
\bibitem [{\citenamefont {Balents}\ and\ \citenamefont
  {Fisher}(1996)}]{PhysRevB.53.12133}%
  \BibitemOpen
  \bibfield  {author} {\bibinfo {author} {\bibfnamefont {L.}~\bibnamefont
  {Balents}}\ and\ \bibinfo {author} {\bibfnamefont {M.~P.~A.}\ \bibnamefont
  {Fisher}},\ }\href {\doibase 10.1103/PhysRevB.53.12133} {\bibfield  {journal}
  {\bibinfo  {journal} {Phys. Rev. B}\ }\textbf {\bibinfo {volume} {53}},\
  \bibinfo {pages} {12133} (\bibinfo {year} {1996})}\BibitemShut {NoStop}%
\bibitem [{Note1()}]{Note1}%
  \BibitemOpen
  \bibinfo {note} {Note that Ref.~\cite {Raghu2EA10} uses $V_e = U + U^2 \chi
  (\protect \hat {k}_1 + \protect \hat {k}_2)$ instead of $V_e = U + U^2 \chi
  (\protect \hat {k}_1 - \protect \hat {k}_2)$, but these two choices are
  equivalent since we only care about integrals of the type $\DOTSI \intop
  \ilimits@ d\protect \hat {k}_2 V_e(\protect \hat {k}_1,\protect \hat {k}_2)
  \Delta (\protect \hat {k}_2)$ with even gap functions ($\Delta (\protect \hat
  {k}_2)=\Delta (-\protect \hat {k}_2)$). It is easy to see that these
  integrals are equal for either choice of $V_e$.}\BibitemShut {Stop}%
\bibitem [{Note2()}]{Note2}%
  \BibitemOpen
  \bibinfo {note} {The $U$ term in the singlet channel gives a zero
  contribution for any $m>0$.}\BibitemShut {Stop}%
\bibitem [{\citenamefont {Wang}\ and\ \citenamefont
  {Fu}(2017)}]{PhysRevLett.119.187003}%
  \BibitemOpen
  \bibfield  {author} {\bibinfo {author} {\bibfnamefont {Y.}~\bibnamefont
  {Wang}}\ and\ \bibinfo {author} {\bibfnamefont {L.}~\bibnamefont {Fu}},\
  }\href {\doibase 10.1103/PhysRevLett.119.187003} {\bibfield  {journal}
  {\bibinfo  {journal} {Phys. Rev. Lett.}\ }\textbf {\bibinfo {volume} {119}},\
  \bibinfo {pages} {187003} (\bibinfo {year} {2017})}\BibitemShut {NoStop}%
\bibitem [{Note3()}]{Note3}%
  \BibitemOpen
  \bibinfo {note} {$\Delta _o$ should actually be described by a vector order
  parameter $\protect \vec {d}$ describing the spin component of the Cooper
  pair\cite {Sigrist05}, but we are free to choose the orientation of $\protect
  \vec {d}$ without loss of generality since the model is $SU(2)$-symmetric. We
  chose $\protect \vec {d} \parallel \protect \hat {z}$ in this section to
  simplify notations.}\BibitemShut {Stop}%
\bibitem [{Note4()}]{Note4}%
  \BibitemOpen
  \bibinfo {note} {We have omitted a term proportional to $\psi _e^4 - \psi
  _o^4$ for simplicity because it does not change the physics
  qualitatively.}\BibitemShut {Stop}%
\bibitem [{\citenamefont {Frank}\ and\ \citenamefont
  {Lemm}(2016)}]{Frank:2016aa}%
  \BibitemOpen
  \bibfield  {author} {\bibinfo {author} {\bibfnamefont {R.~L.}\ \bibnamefont
  {Frank}}\ and\ \bibinfo {author} {\bibfnamefont {M.}~\bibnamefont {Lemm}},\
  }\href {\doibase 10.1007/s00023-016-0473-x} {\bibfield  {journal} {\bibinfo
  {journal} {Annales Henri Poincar{\'e}}\ }\textbf {\bibinfo {volume} {17}},\
  \bibinfo {pages} {2285} (\bibinfo {year} {2016})}\BibitemShut {NoStop}%
\bibitem [{\citenamefont {Sigrist}(2005)}]{Sigrist05}%
  \BibitemOpen
  \bibfield  {author} {\bibinfo {author} {\bibfnamefont {M.}~\bibnamefont
  {Sigrist}},\ }\href {\doibase 10.1063/1.2080350} {\bibfield  {journal}
  {\bibinfo  {journal} {AIP Conf. Proc.}\ }\textbf {\bibinfo {volume} {789}},\
  \bibinfo {pages} {165} (\bibinfo {year} {2005})}\BibitemShut {NoStop}%
\bibitem [{\citenamefont {NishiZaki}\ \emph {et~al.}(2000)\citenamefont
  {NishiZaki}, \citenamefont {Maeno},\ and\ \citenamefont
  {Mao}}]{NishiZakiEA00}%
  \BibitemOpen
  \bibfield  {author} {\bibinfo {author} {\bibfnamefont {S.}~\bibnamefont
  {NishiZaki}}, \bibinfo {author} {\bibfnamefont {Y.}~\bibnamefont {Maeno}}, \
  and\ \bibinfo {author} {\bibfnamefont {Z.}~\bibnamefont {Mao}},\ }\href
  {\doibase 10.1143/JPSJ.69.572} {\bibfield  {journal} {\bibinfo  {journal} {J.
  Phys. Soc. Jpn.}\ }\textbf {\bibinfo {volume} {69}},\ \bibinfo {pages} {572}
  (\bibinfo {year} {2000})}\BibitemShut {NoStop}%
\bibitem [{\citenamefont {Bonalde}\ \emph {et~al.}(2000)\citenamefont
  {Bonalde}, \citenamefont {Yanoff}, \citenamefont {Salamon}, \citenamefont
  {Van~Harlingen}, \citenamefont {Chia}, \citenamefont {Mao},\ and\
  \citenamefont {Maeno}}]{PhysRevLett.85.4775}%
  \BibitemOpen
  \bibfield  {author} {\bibinfo {author} {\bibfnamefont {I.}~\bibnamefont
  {Bonalde}}, \bibinfo {author} {\bibfnamefont {B.~D.}\ \bibnamefont {Yanoff}},
  \bibinfo {author} {\bibfnamefont {M.~B.}\ \bibnamefont {Salamon}}, \bibinfo
  {author} {\bibfnamefont {D.~J.}\ \bibnamefont {Van~Harlingen}}, \bibinfo
  {author} {\bibfnamefont {E.~M.~E.}\ \bibnamefont {Chia}}, \bibinfo {author}
  {\bibfnamefont {Z.~Q.}\ \bibnamefont {Mao}}, \ and\ \bibinfo {author}
  {\bibfnamefont {Y.}~\bibnamefont {Maeno}},\ }\href {\doibase
  10.1103/PhysRevLett.85.4775} {\bibfield  {journal} {\bibinfo  {journal}
  {Phys. Rev. Lett.}\ }\textbf {\bibinfo {volume} {85}},\ \bibinfo {pages}
  {4775} (\bibinfo {year} {2000})}\BibitemShut {NoStop}%
\bibitem [{\citenamefont {Lupien}\ \emph {et~al.}(2001)\citenamefont {Lupien},
  \citenamefont {MacFarlane}, \citenamefont {Proust}, \citenamefont
  {Taillefer}, \citenamefont {Mao},\ and\ \citenamefont {Maeno}}]{LupienEA01}%
  \BibitemOpen
  \bibfield  {author} {\bibinfo {author} {\bibfnamefont {C.}~\bibnamefont
  {Lupien}}, \bibinfo {author} {\bibfnamefont {W.~A.}\ \bibnamefont
  {MacFarlane}}, \bibinfo {author} {\bibfnamefont {C.}~\bibnamefont {Proust}},
  \bibinfo {author} {\bibfnamefont {L.}~\bibnamefont {Taillefer}}, \bibinfo
  {author} {\bibfnamefont {Z.~Q.}\ \bibnamefont {Mao}}, \ and\ \bibinfo
  {author} {\bibfnamefont {Y.}~\bibnamefont {Maeno}},\ }\href {\doibase
  10.1103/PhysRevLett.86.5986} {\bibfield  {journal} {\bibinfo  {journal}
  {Phys. Rev. Lett.}\ }\textbf {\bibinfo {volume} {86}},\ \bibinfo {pages}
  {5986} (\bibinfo {year} {2001})}\BibitemShut {NoStop}%
\bibitem [{\citenamefont {Hassinger}\ \emph {et~al.}(2017)\citenamefont
  {Hassinger}, \citenamefont {Bourgeois-Hope}, \citenamefont {Taniguchi},
  \citenamefont {Ren\'e~de Cotret}, \citenamefont {Grissonnanche},
  \citenamefont {Anwar}, \citenamefont {Maeno}, \citenamefont
  {Doiron-Leyraud},\ and\ \citenamefont {Taillefer}}]{HassingerEA17}%
  \BibitemOpen
  \bibfield  {author} {\bibinfo {author} {\bibfnamefont {E.}~\bibnamefont
  {Hassinger}}, \bibinfo {author} {\bibfnamefont {P.}~\bibnamefont
  {Bourgeois-Hope}}, \bibinfo {author} {\bibfnamefont {H.}~\bibnamefont
  {Taniguchi}}, \bibinfo {author} {\bibfnamefont {S.}~\bibnamefont {Ren\'e~de
  Cotret}}, \bibinfo {author} {\bibfnamefont {G.}~\bibnamefont
  {Grissonnanche}}, \bibinfo {author} {\bibfnamefont {M.~S.}\ \bibnamefont
  {Anwar}}, \bibinfo {author} {\bibfnamefont {Y.}~\bibnamefont {Maeno}},
  \bibinfo {author} {\bibfnamefont {N.}~\bibnamefont {Doiron-Leyraud}}, \ and\
  \bibinfo {author} {\bibfnamefont {L.}~\bibnamefont {Taillefer}},\ }\href
  {\doibase 10.1103/PhysRevX.7.011032} {\bibfield  {journal} {\bibinfo
  {journal} {Phys. Rev. X}\ }\textbf {\bibinfo {volume} {7}},\ \bibinfo {pages}
  {011032} (\bibinfo {year} {2017})}\BibitemShut {NoStop}%
\bibitem [{\citenamefont {Sharma}\ \emph {et~al.}(2020)\citenamefont {Sharma},
  \citenamefont {Edkins}, \citenamefont {Wang}, \citenamefont {Kostin},
  \citenamefont {Sow}, \citenamefont {Maeno}, \citenamefont {Mackenzie},
  \citenamefont {Davis},\ and\ \citenamefont {Madhavan}}]{Sharma5222}%
  \BibitemOpen
  \bibfield  {author} {\bibinfo {author} {\bibfnamefont {R.}~\bibnamefont
  {Sharma}}, \bibinfo {author} {\bibfnamefont {S.~D.}\ \bibnamefont {Edkins}},
  \bibinfo {author} {\bibfnamefont {Z.}~\bibnamefont {Wang}}, \bibinfo {author}
  {\bibfnamefont {A.}~\bibnamefont {Kostin}}, \bibinfo {author} {\bibfnamefont
  {C.}~\bibnamefont {Sow}}, \bibinfo {author} {\bibfnamefont {Y.}~\bibnamefont
  {Maeno}}, \bibinfo {author} {\bibfnamefont {A.~P.}\ \bibnamefont
  {Mackenzie}}, \bibinfo {author} {\bibfnamefont {J.~C.~S.}\ \bibnamefont
  {Davis}}, \ and\ \bibinfo {author} {\bibfnamefont {V.}~\bibnamefont
  {Madhavan}},\ }\href {\doibase 10.1073/pnas.1916463117} {\bibfield  {journal}
  {\bibinfo  {journal} {Proceedings of the National Academy of Sciences}\
  }\textbf {\bibinfo {volume} {117}},\ \bibinfo {pages} {5222} (\bibinfo {year}
  {2020})},\ \Eprint
  {http://arxiv.org/abs/https://www.pnas.org/content/117/10/5222.full.pdf}
  {https://www.pnas.org/content/117/10/5222.full.pdf} \BibitemShut {NoStop}%
\bibitem [{\citenamefont {Hicks}\ \emph {et~al.}(2010)\citenamefont {Hicks},
  \citenamefont {Kirtley}, \citenamefont {Lippman}, \citenamefont {Koshnick},
  \citenamefont {Huber}, \citenamefont {Maeno}, \citenamefont {Yuhasz},
  \citenamefont {Maple},\ and\ \citenamefont {Moler}}]{PhysRevB.81.214501}%
  \BibitemOpen
  \bibfield  {author} {\bibinfo {author} {\bibfnamefont {C.~W.}\ \bibnamefont
  {Hicks}}, \bibinfo {author} {\bibfnamefont {J.~R.}\ \bibnamefont {Kirtley}},
  \bibinfo {author} {\bibfnamefont {T.~M.}\ \bibnamefont {Lippman}}, \bibinfo
  {author} {\bibfnamefont {N.~C.}\ \bibnamefont {Koshnick}}, \bibinfo {author}
  {\bibfnamefont {M.~E.}\ \bibnamefont {Huber}}, \bibinfo {author}
  {\bibfnamefont {Y.}~\bibnamefont {Maeno}}, \bibinfo {author} {\bibfnamefont
  {W.~M.}\ \bibnamefont {Yuhasz}}, \bibinfo {author} {\bibfnamefont {M.~B.}\
  \bibnamefont {Maple}}, \ and\ \bibinfo {author} {\bibfnamefont {K.~A.}\
  \bibnamefont {Moler}},\ }\href {\doibase 10.1103/PhysRevB.81.214501}
  {\bibfield  {journal} {\bibinfo  {journal} {Phys. Rev. B}\ }\textbf {\bibinfo
  {volume} {81}},\ \bibinfo {pages} {214501} (\bibinfo {year}
  {2010})}\BibitemShut {NoStop}%
\bibitem [{\citenamefont {Huang}\ \emph {et~al.}(2015)\citenamefont {Huang},
  \citenamefont {Lederer}, \citenamefont {Taylor},\ and\ \citenamefont
  {Kallin}}]{PhysRevB.91.094507}%
  \BibitemOpen
  \bibfield  {author} {\bibinfo {author} {\bibfnamefont {W.}~\bibnamefont
  {Huang}}, \bibinfo {author} {\bibfnamefont {S.}~\bibnamefont {Lederer}},
  \bibinfo {author} {\bibfnamefont {E.}~\bibnamefont {Taylor}}, \ and\ \bibinfo
  {author} {\bibfnamefont {C.}~\bibnamefont {Kallin}},\ }\href {\doibase
  10.1103/PhysRevB.91.094507} {\bibfield  {journal} {\bibinfo  {journal} {Phys.
  Rev. B}\ }\textbf {\bibinfo {volume} {91}},\ \bibinfo {pages} {094507}
  (\bibinfo {year} {2015})}\BibitemShut {NoStop}%
\bibitem [{\citenamefont {Huang}\ \emph {et~al.}(2014)\citenamefont {Huang},
  \citenamefont {Taylor},\ and\ \citenamefont {Kallin}}]{PhysRevB.90.224519}%
  \BibitemOpen
  \bibfield  {author} {\bibinfo {author} {\bibfnamefont {W.}~\bibnamefont
  {Huang}}, \bibinfo {author} {\bibfnamefont {E.}~\bibnamefont {Taylor}}, \
  and\ \bibinfo {author} {\bibfnamefont {C.}~\bibnamefont {Kallin}},\ }\href
  {\doibase 10.1103/PhysRevB.90.224519} {\bibfield  {journal} {\bibinfo
  {journal} {Phys. Rev. B}\ }\textbf {\bibinfo {volume} {90}},\ \bibinfo
  {pages} {224519} (\bibinfo {year} {2014})}\BibitemShut {NoStop}%
\bibitem [{\citenamefont {Lederer}\ \emph {et~al.}(2014)\citenamefont
  {Lederer}, \citenamefont {Huang}, \citenamefont {Taylor}, \citenamefont
  {Raghu},\ and\ \citenamefont {Kallin}}]{PhysRevB.90.134521}%
  \BibitemOpen
  \bibfield  {author} {\bibinfo {author} {\bibfnamefont {S.}~\bibnamefont
  {Lederer}}, \bibinfo {author} {\bibfnamefont {W.}~\bibnamefont {Huang}},
  \bibinfo {author} {\bibfnamefont {E.}~\bibnamefont {Taylor}}, \bibinfo
  {author} {\bibfnamefont {S.}~\bibnamefont {Raghu}}, \ and\ \bibinfo {author}
  {\bibfnamefont {C.}~\bibnamefont {Kallin}},\ }\href {\doibase
  10.1103/PhysRevB.90.134521} {\bibfield  {journal} {\bibinfo  {journal} {Phys.
  Rev. B}\ }\textbf {\bibinfo {volume} {90}},\ \bibinfo {pages} {134521}
  (\bibinfo {year} {2014})}\BibitemShut {NoStop}%
\bibitem [{\citenamefont {Scaffidi}\ and\ \citenamefont
  {Simon}(2015)}]{ScaffidiSimon15}%
  \BibitemOpen
  \bibfield  {author} {\bibinfo {author} {\bibfnamefont {T.}~\bibnamefont
  {Scaffidi}}\ and\ \bibinfo {author} {\bibfnamefont {S.~H.}\ \bibnamefont
  {Simon}},\ }\href {\doibase 10.1103/PhysRevLett.115.087003} {\bibfield
  {journal} {\bibinfo  {journal} {Phys. Rev. Lett.}\ }\textbf {\bibinfo
  {volume} {115}},\ \bibinfo {pages} {087003} (\bibinfo {year}
  {2015})}\BibitemShut {NoStop}%
\bibitem [{Note5()}]{Note5}%
  \BibitemOpen
  \bibinfo {note} {These gradient terms become important close to sample edges,
  domain walls, or defects, in the vicinity of which the order parameter is not
  spatially homogeneous.}\BibitemShut {Stop}%
\bibitem [{\citenamefont {Achermann}\ \emph {et~al.}(2014)\citenamefont
  {Achermann}, \citenamefont {Neupert}, \citenamefont {Arahata},\ and\
  \citenamefont {Sigrist}}]{doi:10.7566/JPSJ.83.044712}%
  \BibitemOpen
  \bibfield  {author} {\bibinfo {author} {\bibfnamefont {M.}~\bibnamefont
  {Achermann}}, \bibinfo {author} {\bibfnamefont {T.}~\bibnamefont {Neupert}},
  \bibinfo {author} {\bibfnamefont {E.}~\bibnamefont {Arahata}}, \ and\
  \bibinfo {author} {\bibfnamefont {M.}~\bibnamefont {Sigrist}},\ }\href
  {\doibase 10.7566/JPSJ.83.044712} {\bibfield  {journal} {\bibinfo  {journal}
  {Journal of the Physical Society of Japan}\ }\textbf {\bibinfo {volume}
  {83}},\ \bibinfo {pages} {044712} (\bibinfo {year} {2014})},\ \Eprint
  {http://arxiv.org/abs/https://doi.org/10.7566/JPSJ.83.044712}
  {https://doi.org/10.7566/JPSJ.83.044712} \BibitemShut {NoStop}%
\bibitem [{\citenamefont {{Yang}}\ \emph {et~al.}(2017)\citenamefont {{Yang}},
  \citenamefont {{Xu}},\ and\ \citenamefont {{Wu}}}]{2017arXiv171105241Y}%
  \BibitemOpen
  \bibfield  {author} {\bibinfo {author} {\bibfnamefont {W.}~\bibnamefont
  {{Yang}}}, \bibinfo {author} {\bibfnamefont {C.}~\bibnamefont {{Xu}}}, \ and\
  \bibinfo {author} {\bibfnamefont {C.}~\bibnamefont {{Wu}}},\ }\href@noop {}
  {\bibfield  {journal} {\bibinfo  {journal} {arXiv e-prints}\ ,\ \bibinfo
  {eid} {arXiv:1711.05241}} (\bibinfo {year} {2017})},\ \Eprint
  {http://arxiv.org/abs/1711.05241} {arXiv:1711.05241 [cond-mat.supr-con]}
  \BibitemShut {NoStop}%
\bibitem [{\citenamefont {Robins}\ and\ \citenamefont
  {Brydon}(2018)}]{Robins_2018}%
  \BibitemOpen
  \bibfield  {author} {\bibinfo {author} {\bibfnamefont {A.}~\bibnamefont
  {Robins}}\ and\ \bibinfo {author} {\bibfnamefont {P.}~\bibnamefont
  {Brydon}},\ }\href {\doibase 10.1088/1361-648x/aade6a} {\bibfield  {journal}
  {\bibinfo  {journal} {Journal of Physics: Condensed Matter}\ }\textbf
  {\bibinfo {volume} {30}},\ \bibinfo {pages} {405602} (\bibinfo {year}
  {2018})}\BibitemShut {NoStop}%
\bibitem [{\citenamefont {Hicks}\ \emph {et~al.}(2014)\citenamefont {Hicks},
  \citenamefont {Brodsky}, \citenamefont {Yelland}, \citenamefont {Gibbs},
  \citenamefont {Bruin}, \citenamefont {Barber}, \citenamefont {Edkins},
  \citenamefont {Nishimura}, \citenamefont {Yonezawa}, \citenamefont {Maeno},\
  and\ \citenamefont {Mackenzie}}]{HicksEA14}%
  \BibitemOpen
  \bibfield  {author} {\bibinfo {author} {\bibfnamefont {C.~W.}\ \bibnamefont
  {Hicks}}, \bibinfo {author} {\bibfnamefont {D.~O.}\ \bibnamefont {Brodsky}},
  \bibinfo {author} {\bibfnamefont {E.~A.}\ \bibnamefont {Yelland}}, \bibinfo
  {author} {\bibfnamefont {A.~S.}\ \bibnamefont {Gibbs}}, \bibinfo {author}
  {\bibfnamefont {J.~A.~N.}\ \bibnamefont {Bruin}}, \bibinfo {author}
  {\bibfnamefont {M.~E.}\ \bibnamefont {Barber}}, \bibinfo {author}
  {\bibfnamefont {S.~D.}\ \bibnamefont {Edkins}}, \bibinfo {author}
  {\bibfnamefont {K.}~\bibnamefont {Nishimura}}, \bibinfo {author}
  {\bibfnamefont {S.}~\bibnamefont {Yonezawa}}, \bibinfo {author}
  {\bibfnamefont {Y.}~\bibnamefont {Maeno}}, \ and\ \bibinfo {author}
  {\bibfnamefont {A.~P.}\ \bibnamefont {Mackenzie}},\ }\href {\doibase
  10.1126/science.1248292} {\bibfield  {journal} {\bibinfo  {journal}
  {Science}\ }\textbf {\bibinfo {volume} {344}},\ \bibinfo {pages} {283}
  (\bibinfo {year} {2014})}\BibitemShut {NoStop}%
\bibitem [{\citenamefont {Steppke}\ \emph {et~al.}(2017)\citenamefont
  {Steppke}, \citenamefont {Zhao}, \citenamefont {Barber}, \citenamefont
  {Scaffidi}, \citenamefont {Jerzembeck}, \citenamefont {Rosner}, \citenamefont
  {Gibbs}, \citenamefont {Maeno}, \citenamefont {Simon}, \citenamefont
  {Mackenzie},\ and\ \citenamefont {Hicks}}]{SteppkeEA17}%
  \BibitemOpen
  \bibfield  {author} {\bibinfo {author} {\bibfnamefont {A.}~\bibnamefont
  {Steppke}}, \bibinfo {author} {\bibfnamefont {L.}~\bibnamefont {Zhao}},
  \bibinfo {author} {\bibfnamefont {M.~E.}\ \bibnamefont {Barber}}, \bibinfo
  {author} {\bibfnamefont {T.}~\bibnamefont {Scaffidi}}, \bibinfo {author}
  {\bibfnamefont {F.}~\bibnamefont {Jerzembeck}}, \bibinfo {author}
  {\bibfnamefont {H.}~\bibnamefont {Rosner}}, \bibinfo {author} {\bibfnamefont
  {A.~S.}\ \bibnamefont {Gibbs}}, \bibinfo {author} {\bibfnamefont
  {Y.}~\bibnamefont {Maeno}}, \bibinfo {author} {\bibfnamefont {S.~H.}\
  \bibnamefont {Simon}}, \bibinfo {author} {\bibfnamefont {A.~P.}\ \bibnamefont
  {Mackenzie}}, \ and\ \bibinfo {author} {\bibfnamefont {C.~W.}\ \bibnamefont
  {Hicks}},\ }\href {\doibase 10.1126/science.aaf9398} {\bibfield  {journal}
  {\bibinfo  {journal} {Science}\ }\textbf {\bibinfo {volume} {355}} (\bibinfo
  {year} {2017}),\ 10.1126/science.aaf9398}\BibitemShut {NoStop}%
\bibitem [{\citenamefont {Nelson}\ \emph {et~al.}(2004)\citenamefont {Nelson},
  \citenamefont {Mao}, \citenamefont {Maeno},\ and\ \citenamefont
  {Liu}}]{Nelson1151}%
  \BibitemOpen
  \bibfield  {author} {\bibinfo {author} {\bibfnamefont {K.~D.}\ \bibnamefont
  {Nelson}}, \bibinfo {author} {\bibfnamefont {Z.~Q.}\ \bibnamefont {Mao}},
  \bibinfo {author} {\bibfnamefont {Y.}~\bibnamefont {Maeno}}, \ and\ \bibinfo
  {author} {\bibfnamefont {Y.}~\bibnamefont {Liu}},\ }\href {\doibase
  10.1126/science.1103881} {\bibfield  {journal} {\bibinfo  {journal}
  {Science}\ }\textbf {\bibinfo {volume} {306}},\ \bibinfo {pages} {1151}
  (\bibinfo {year} {2004})},\ \Eprint
  {http://arxiv.org/abs/https://science.sciencemag.org/content/306/5699/1151.full.pdf}
  {https://science.sciencemag.org/content/306/5699/1151.full.pdf} \BibitemShut
  {NoStop}%
\bibitem [{\citenamefont {Kidwingira}\ \emph {et~al.}(2006)\citenamefont
  {Kidwingira}, \citenamefont {Strand}, \citenamefont {Van~Harlingen},\ and\
  \citenamefont {Maeno}}]{Kidwingira1267}%
  \BibitemOpen
  \bibfield  {author} {\bibinfo {author} {\bibfnamefont {F.}~\bibnamefont
  {Kidwingira}}, \bibinfo {author} {\bibfnamefont {J.~D.}\ \bibnamefont
  {Strand}}, \bibinfo {author} {\bibfnamefont {D.~J.}\ \bibnamefont
  {Van~Harlingen}}, \ and\ \bibinfo {author} {\bibfnamefont {Y.}~\bibnamefont
  {Maeno}},\ }\href {\doibase 10.1126/science.1133239} {\bibfield  {journal}
  {\bibinfo  {journal} {Science}\ }\textbf {\bibinfo {volume} {314}},\ \bibinfo
  {pages} {1267} (\bibinfo {year} {2006})},\ \Eprint
  {http://arxiv.org/abs/https://science.sciencemag.org/content/314/5803/1267.full.pdf}
  {https://science.sciencemag.org/content/314/5803/1267.full.pdf} \BibitemShut
  {NoStop}%
\bibitem [{\citenamefont {Anwar}\ \emph {et~al.}(2017)\citenamefont {Anwar},
  \citenamefont {Ishiguro}, \citenamefont {Nakamura}, \citenamefont {Yakabe},
  \citenamefont {Yonezawa}, \citenamefont {Takayanagi},\ and\ \citenamefont
  {Maeno}}]{PhysRevB.95.224509}%
  \BibitemOpen
  \bibfield  {author} {\bibinfo {author} {\bibfnamefont {M.~S.}\ \bibnamefont
  {Anwar}}, \bibinfo {author} {\bibfnamefont {R.}~\bibnamefont {Ishiguro}},
  \bibinfo {author} {\bibfnamefont {T.}~\bibnamefont {Nakamura}}, \bibinfo
  {author} {\bibfnamefont {M.}~\bibnamefont {Yakabe}}, \bibinfo {author}
  {\bibfnamefont {S.}~\bibnamefont {Yonezawa}}, \bibinfo {author}
  {\bibfnamefont {H.}~\bibnamefont {Takayanagi}}, \ and\ \bibinfo {author}
  {\bibfnamefont {Y.}~\bibnamefont {Maeno}},\ }\href {\doibase
  10.1103/PhysRevB.95.224509} {\bibfield  {journal} {\bibinfo  {journal} {Phys.
  Rev. B}\ }\textbf {\bibinfo {volume} {95}},\ \bibinfo {pages} {224509}
  (\bibinfo {year} {2017})}\BibitemShut {NoStop}%
\bibitem [{\citenamefont {Jang}\ \emph {et~al.}(2011)\citenamefont {Jang},
  \citenamefont {Ferguson}, \citenamefont {Vakaryuk}, \citenamefont {Budakian},
  \citenamefont {Chung}, \citenamefont {Goldbart},\ and\ \citenamefont
  {Maeno}}]{Jang186}%
  \BibitemOpen
  \bibfield  {author} {\bibinfo {author} {\bibfnamefont {J.}~\bibnamefont
  {Jang}}, \bibinfo {author} {\bibfnamefont {D.~G.}\ \bibnamefont {Ferguson}},
  \bibinfo {author} {\bibfnamefont {V.}~\bibnamefont {Vakaryuk}}, \bibinfo
  {author} {\bibfnamefont {R.}~\bibnamefont {Budakian}}, \bibinfo {author}
  {\bibfnamefont {S.~B.}\ \bibnamefont {Chung}}, \bibinfo {author}
  {\bibfnamefont {P.~M.}\ \bibnamefont {Goldbart}}, \ and\ \bibinfo {author}
  {\bibfnamefont {Y.}~\bibnamefont {Maeno}},\ }\href {\doibase
  10.1126/science.1193839} {\bibfield  {journal} {\bibinfo  {journal}
  {Science}\ }\textbf {\bibinfo {volume} {331}},\ \bibinfo {pages} {186}
  (\bibinfo {year} {2011})},\ \Eprint
  {http://arxiv.org/abs/https://science.sciencemag.org/content/331/6014/186.full.pdf}
  {https://science.sciencemag.org/content/331/6014/186.full.pdf} \BibitemShut
  {NoStop}%
\bibitem [{\citenamefont {Anwar}\ \emph {et~al.}(2019)\citenamefont {Anwar},
  \citenamefont {Kunieda}, \citenamefont {Ishiguro}, \citenamefont {Lee},
  \citenamefont {Olthof}, \citenamefont {Robinson}, \citenamefont {Yonezawa},
  \citenamefont {Noh},\ and\ \citenamefont {Maeno}}]{PhysRevB.100.024516}%
  \BibitemOpen
  \bibfield  {author} {\bibinfo {author} {\bibfnamefont {M.~S.}\ \bibnamefont
  {Anwar}}, \bibinfo {author} {\bibfnamefont {M.}~\bibnamefont {Kunieda}},
  \bibinfo {author} {\bibfnamefont {R.}~\bibnamefont {Ishiguro}}, \bibinfo
  {author} {\bibfnamefont {S.~R.}\ \bibnamefont {Lee}}, \bibinfo {author}
  {\bibfnamefont {L.~A. B.~O.}\ \bibnamefont {Olthof}}, \bibinfo {author}
  {\bibfnamefont {J.~W.~A.}\ \bibnamefont {Robinson}}, \bibinfo {author}
  {\bibfnamefont {S.}~\bibnamefont {Yonezawa}}, \bibinfo {author}
  {\bibfnamefont {T.~W.}\ \bibnamefont {Noh}}, \ and\ \bibinfo {author}
  {\bibfnamefont {Y.}~\bibnamefont {Maeno}},\ }\href {\doibase
  10.1103/PhysRevB.100.024516} {\bibfield  {journal} {\bibinfo  {journal}
  {Phys. Rev. B}\ }\textbf {\bibinfo {volume} {100}},\ \bibinfo {pages}
  {024516} (\bibinfo {year} {2019})}\BibitemShut {NoStop}%
\bibitem [{\citenamefont {Wang}\ \emph {et~al.}(2020)\citenamefont {Wang},
  \citenamefont {Wang},\ and\ \citenamefont {Kallin}}]{PhysRevB.101.064507}%
  \BibitemOpen
  \bibfield  {author} {\bibinfo {author} {\bibfnamefont {Z.}~\bibnamefont
  {Wang}}, \bibinfo {author} {\bibfnamefont {X.}~\bibnamefont {Wang}}, \ and\
  \bibinfo {author} {\bibfnamefont {C.}~\bibnamefont {Kallin}},\ }\href
  {\doibase 10.1103/PhysRevB.101.064507} {\bibfield  {journal} {\bibinfo
  {journal} {Phys. Rev. B}\ }\textbf {\bibinfo {volume} {101}},\ \bibinfo
  {pages} {064507} (\bibinfo {year} {2020})}\BibitemShut {NoStop}%
\bibitem [{\citenamefont {Taylor}\ and\ \citenamefont
  {Kallin}(2012)}]{PhysRevLett.108.157001}%
  \BibitemOpen
  \bibfield  {author} {\bibinfo {author} {\bibfnamefont {E.}~\bibnamefont
  {Taylor}}\ and\ \bibinfo {author} {\bibfnamefont {C.}~\bibnamefont
  {Kallin}},\ }\href {\doibase 10.1103/PhysRevLett.108.157001} {\bibfield
  {journal} {\bibinfo  {journal} {Phys. Rev. Lett.}\ }\textbf {\bibinfo
  {volume} {108}},\ \bibinfo {pages} {157001} (\bibinfo {year}
  {2012})}\BibitemShut {NoStop}%
\bibitem [{\citenamefont {Kapitulnik}\ \emph {et~al.}(2009)\citenamefont
  {Kapitulnik}, \citenamefont {Xia}, \citenamefont {Schemm},\ and\
  \citenamefont {Palevski}}]{Kapitulnik_2009}%
  \BibitemOpen
  \bibfield  {author} {\bibinfo {author} {\bibfnamefont {A.}~\bibnamefont
  {Kapitulnik}}, \bibinfo {author} {\bibfnamefont {J.}~\bibnamefont {Xia}},
  \bibinfo {author} {\bibfnamefont {E.}~\bibnamefont {Schemm}}, \ and\ \bibinfo
  {author} {\bibfnamefont {A.}~\bibnamefont {Palevski}},\ }\href {\doibase
  10.1088/1367-2630/11/5/055060} {\bibfield  {journal} {\bibinfo  {journal}
  {New Journal of Physics}\ }\textbf {\bibinfo {volume} {11}},\ \bibinfo
  {pages} {055060} (\bibinfo {year} {2009})}\BibitemShut {NoStop}%
\bibitem [{\citenamefont {Haverkort}\ \emph {et~al.}(2008)\citenamefont
  {Haverkort}, \citenamefont {Elfimov}, \citenamefont {Tjeng}, \citenamefont
  {Sawatzky},\ and\ \citenamefont {Damascelli}}]{Damascelli08}%
  \BibitemOpen
  \bibfield  {author} {\bibinfo {author} {\bibfnamefont {M.~W.}\ \bibnamefont
  {Haverkort}}, \bibinfo {author} {\bibfnamefont {I.~S.}\ \bibnamefont
  {Elfimov}}, \bibinfo {author} {\bibfnamefont {L.~H.}\ \bibnamefont {Tjeng}},
  \bibinfo {author} {\bibfnamefont {G.~A.}\ \bibnamefont {Sawatzky}}, \ and\
  \bibinfo {author} {\bibfnamefont {A.}~\bibnamefont {Damascelli}},\ }\href
  {\doibase 10.1103/PhysRevLett.101.026406} {\bibfield  {journal} {\bibinfo
  {journal} {Phys. Rev. Lett.}\ }\textbf {\bibinfo {volume} {101}},\ \bibinfo
  {pages} {026406} (\bibinfo {year} {2008})}\BibitemShut {NoStop}%
\bibitem [{\citenamefont {Wang}\ \emph {et~al.}(2013)\citenamefont {Wang},
  \citenamefont {Platt}, \citenamefont {Yang}, \citenamefont {Honerkamp},
  \citenamefont {Zhang}, \citenamefont {Hanke}, \citenamefont {Rice},\ and\
  \citenamefont {Thomale}}]{Wang_2013}%
  \BibitemOpen
  \bibfield  {author} {\bibinfo {author} {\bibfnamefont {Q.~H.}\ \bibnamefont
  {Wang}}, \bibinfo {author} {\bibfnamefont {C.}~\bibnamefont {Platt}},
  \bibinfo {author} {\bibfnamefont {Y.}~\bibnamefont {Yang}}, \bibinfo {author}
  {\bibfnamefont {C.}~\bibnamefont {Honerkamp}}, \bibinfo {author}
  {\bibfnamefont {F.~C.}\ \bibnamefont {Zhang}}, \bibinfo {author}
  {\bibfnamefont {W.}~\bibnamefont {Hanke}}, \bibinfo {author} {\bibfnamefont
  {T.~M.}\ \bibnamefont {Rice}}, \ and\ \bibinfo {author} {\bibfnamefont
  {R.}~\bibnamefont {Thomale}},\ }\href {\doibase 10.1209/0295-5075/104/17013}
  {\bibfield  {journal} {\bibinfo  {journal} {{EPL} (Europhysics Letters)}\
  }\textbf {\bibinfo {volume} {104}},\ \bibinfo {pages} {17013} (\bibinfo
  {year} {2013})}\BibitemShut {NoStop}%
\bibitem [{\citenamefont {Huo}\ \emph {et~al.}(2013)\citenamefont {Huo},
  \citenamefont {Rice},\ and\ \citenamefont {Zhang}}]{HuoEA13}%
  \BibitemOpen
  \bibfield  {author} {\bibinfo {author} {\bibfnamefont {J.-W.}\ \bibnamefont
  {Huo}}, \bibinfo {author} {\bibfnamefont {T.~M.}\ \bibnamefont {Rice}}, \
  and\ \bibinfo {author} {\bibfnamefont {F.-C.}\ \bibnamefont {Zhang}},\ }\href
  {\doibase 10.1103/PhysRevLett.110.167003} {\bibfield  {journal} {\bibinfo
  {journal} {Phys. Rev. Lett.}\ }\textbf {\bibinfo {volume} {110}},\ \bibinfo
  {pages} {167003} (\bibinfo {year} {2013})}\BibitemShut {NoStop}%
\bibitem [{\citenamefont {Veenstra}\ \emph {et~al.}(2014)\citenamefont
  {Veenstra}, \citenamefont {Zhu}, \citenamefont {Raichle}, \citenamefont
  {Ludbrook}, \citenamefont {Nicolaou}, \citenamefont {Slomski}, \citenamefont
  {Landolt}, \citenamefont {Kittaka}, \citenamefont {Maeno}, \citenamefont
  {Dil}, \citenamefont {Elfimov}, \citenamefont {Haverkort},\ and\
  \citenamefont {Damascelli}}]{DamascelliEA14}%
  \BibitemOpen
  \bibfield  {author} {\bibinfo {author} {\bibfnamefont {C.~N.}\ \bibnamefont
  {Veenstra}}, \bibinfo {author} {\bibfnamefont {Z.-H.}\ \bibnamefont {Zhu}},
  \bibinfo {author} {\bibfnamefont {M.}~\bibnamefont {Raichle}}, \bibinfo
  {author} {\bibfnamefont {B.~M.}\ \bibnamefont {Ludbrook}}, \bibinfo {author}
  {\bibfnamefont {A.}~\bibnamefont {Nicolaou}}, \bibinfo {author}
  {\bibfnamefont {B.}~\bibnamefont {Slomski}}, \bibinfo {author} {\bibfnamefont
  {G.}~\bibnamefont {Landolt}}, \bibinfo {author} {\bibfnamefont
  {S.}~\bibnamefont {Kittaka}}, \bibinfo {author} {\bibfnamefont
  {Y.}~\bibnamefont {Maeno}}, \bibinfo {author} {\bibfnamefont {J.~H.}\
  \bibnamefont {Dil}}, \bibinfo {author} {\bibfnamefont {I.~S.}\ \bibnamefont
  {Elfimov}}, \bibinfo {author} {\bibfnamefont {M.~W.}\ \bibnamefont
  {Haverkort}}, \ and\ \bibinfo {author} {\bibfnamefont {A.}~\bibnamefont
  {Damascelli}},\ }\href {\doibase 10.1103/PhysRevLett.112.127002} {\bibfield
  {journal} {\bibinfo  {journal} {Phys. Rev. Lett.}\ }\textbf {\bibinfo
  {volume} {112}},\ \bibinfo {pages} {127002} (\bibinfo {year}
  {2014})}\BibitemShut {NoStop}%
\bibitem [{\citenamefont {R\o{}ising}\ \emph {et~al.}(2019)\citenamefont
  {R\o{}ising}, \citenamefont {Scaffidi}, \citenamefont {Flicker},
  \citenamefont {Lange},\ and\ \citenamefont
  {Simon}}]{PhysRevResearch.1.033108}%
  \BibitemOpen
  \bibfield  {author} {\bibinfo {author} {\bibfnamefont {H.~S.}\ \bibnamefont
  {R\o{}ising}}, \bibinfo {author} {\bibfnamefont {T.}~\bibnamefont
  {Scaffidi}}, \bibinfo {author} {\bibfnamefont {F.}~\bibnamefont {Flicker}},
  \bibinfo {author} {\bibfnamefont {G.~F.}\ \bibnamefont {Lange}}, \ and\
  \bibinfo {author} {\bibfnamefont {S.~H.}\ \bibnamefont {Simon}},\ }\href
  {\doibase 10.1103/PhysRevResearch.1.033108} {\bibfield  {journal} {\bibinfo
  {journal} {Phys. Rev. Research}\ }\textbf {\bibinfo {volume} {1}},\ \bibinfo
  {pages} {033108} (\bibinfo {year} {2019})}\BibitemShut {NoStop}%
\bibitem [{\citenamefont {Bourbonnais}\ and\ \citenamefont
  {J{\'e}rome}(2008)}]{bourbonnais2008physics}%
  \BibitemOpen
  \bibfield  {author} {\bibinfo {author} {\bibfnamefont {C.}~\bibnamefont
  {Bourbonnais}}\ and\ \bibinfo {author} {\bibfnamefont {D.}~\bibnamefont
  {J{\'e}rome}},\ }\href@noop {} {\enquote {\bibinfo {title} {Physics of
  organic superconductors and conductors, springerseries in material science
  vol. 110},}\ } (\bibinfo {year} {2008})\BibitemShut {NoStop}%
\bibitem [{\citenamefont {Doiron-Leyraud}\ \emph {et~al.}(2009)\citenamefont
  {Doiron-Leyraud}, \citenamefont {Auban-Senzier}, \citenamefont {Ren\'e~de
  Cotret}, \citenamefont {Bourbonnais}, \citenamefont {J\'erome}, \citenamefont
  {Bechgaard},\ and\ \citenamefont {Taillefer}}]{PhysRevB.80.214531}%
  \BibitemOpen
  \bibfield  {author} {\bibinfo {author} {\bibfnamefont {N.}~\bibnamefont
  {Doiron-Leyraud}}, \bibinfo {author} {\bibfnamefont {P.}~\bibnamefont
  {Auban-Senzier}}, \bibinfo {author} {\bibfnamefont {S.}~\bibnamefont
  {Ren\'e~de Cotret}}, \bibinfo {author} {\bibfnamefont {C.}~\bibnamefont
  {Bourbonnais}}, \bibinfo {author} {\bibfnamefont {D.}~\bibnamefont
  {J\'erome}}, \bibinfo {author} {\bibfnamefont {K.}~\bibnamefont {Bechgaard}},
  \ and\ \bibinfo {author} {\bibfnamefont {L.}~\bibnamefont {Taillefer}},\
  }\href {\doibase 10.1103/PhysRevB.80.214531} {\bibfield  {journal} {\bibinfo
  {journal} {Phys. Rev. B}\ }\textbf {\bibinfo {volume} {80}},\ \bibinfo
  {pages} {214531} (\bibinfo {year} {2009})}\BibitemShut {NoStop}%
\bibitem [{\citenamefont {Cho}\ \emph {et~al.}(2015)\citenamefont {Cho},
  \citenamefont {Platt}, \citenamefont {McKenzie},\ and\ \citenamefont
  {Raghu}}]{WeejeeEA15}%
  \BibitemOpen
  \bibfield  {author} {\bibinfo {author} {\bibfnamefont {W.}~\bibnamefont
  {Cho}}, \bibinfo {author} {\bibfnamefont {C.}~\bibnamefont {Platt}}, \bibinfo
  {author} {\bibfnamefont {R.~H.}\ \bibnamefont {McKenzie}}, \ and\ \bibinfo
  {author} {\bibfnamefont {S.}~\bibnamefont {Raghu}},\ }\href {\doibase
  10.1103/PhysRevB.92.134514} {\bibfield  {journal} {\bibinfo  {journal} {Phys.
  Rev. B}\ }\textbf {\bibinfo {volume} {92}},\ \bibinfo {pages} {134514}
  (\bibinfo {year} {2015})}\BibitemShut {NoStop}%
\bibitem [{\citenamefont {Shigeta}\ \emph {et~al.}(2011)\citenamefont
  {Shigeta}, \citenamefont {Tanaka}, \citenamefont {Kuroki}, \citenamefont
  {Onari},\ and\ \citenamefont {Aizawa}}]{PhysRevB.83.140509}%
  \BibitemOpen
  \bibfield  {author} {\bibinfo {author} {\bibfnamefont {K.}~\bibnamefont
  {Shigeta}}, \bibinfo {author} {\bibfnamefont {Y.}~\bibnamefont {Tanaka}},
  \bibinfo {author} {\bibfnamefont {K.}~\bibnamefont {Kuroki}}, \bibinfo
  {author} {\bibfnamefont {S.}~\bibnamefont {Onari}}, \ and\ \bibinfo {author}
  {\bibfnamefont {H.}~\bibnamefont {Aizawa}},\ }\href {\doibase
  10.1103/PhysRevB.83.140509} {\bibfield  {journal} {\bibinfo  {journal} {Phys.
  Rev. B}\ }\textbf {\bibinfo {volume} {83}},\ \bibinfo {pages} {140509}
  (\bibinfo {year} {2011})}\BibitemShut {NoStop}%
\bibitem [{\citenamefont {Aizawa}\ \emph {et~al.}(2009)\citenamefont {Aizawa},
  \citenamefont {Kuroki}, \citenamefont {Yokoyama},\ and\ \citenamefont
  {Tanaka}}]{PhysRevLett.102.016403}%
  \BibitemOpen
  \bibfield  {author} {\bibinfo {author} {\bibfnamefont {H.}~\bibnamefont
  {Aizawa}}, \bibinfo {author} {\bibfnamefont {K.}~\bibnamefont {Kuroki}},
  \bibinfo {author} {\bibfnamefont {T.}~\bibnamefont {Yokoyama}}, \ and\
  \bibinfo {author} {\bibfnamefont {Y.}~\bibnamefont {Tanaka}},\ }\href
  {\doibase 10.1103/PhysRevLett.102.016403} {\bibfield  {journal} {\bibinfo
  {journal} {Phys. Rev. Lett.}\ }\textbf {\bibinfo {volume} {102}},\ \bibinfo
  {pages} {016403} (\bibinfo {year} {2009})}\BibitemShut {NoStop}%
\bibitem [{\citenamefont {Tanaka}\ and\ \citenamefont
  {Kuroki}(2004)}]{PhysRevB.70.060502}%
  \BibitemOpen
  \bibfield  {author} {\bibinfo {author} {\bibfnamefont {Y.}~\bibnamefont
  {Tanaka}}\ and\ \bibinfo {author} {\bibfnamefont {K.}~\bibnamefont
  {Kuroki}},\ }\href {\doibase 10.1103/PhysRevB.70.060502} {\bibfield
  {journal} {\bibinfo  {journal} {Phys. Rev. B}\ }\textbf {\bibinfo {volume}
  {70}},\ \bibinfo {pages} {060502} (\bibinfo {year} {2004})}\BibitemShut
  {NoStop}%
\bibitem [{\citenamefont {Shigeta}\ \emph {et~al.}(2009)\citenamefont
  {Shigeta}, \citenamefont {Onari}, \citenamefont {Yada},\ and\ \citenamefont
  {Tanaka}}]{PhysRevB.79.174507}%
  \BibitemOpen
  \bibfield  {author} {\bibinfo {author} {\bibfnamefont {K.}~\bibnamefont
  {Shigeta}}, \bibinfo {author} {\bibfnamefont {S.}~\bibnamefont {Onari}},
  \bibinfo {author} {\bibfnamefont {K.}~\bibnamefont {Yada}}, \ and\ \bibinfo
  {author} {\bibfnamefont {Y.}~\bibnamefont {Tanaka}},\ }\href {\doibase
  10.1103/PhysRevB.79.174507} {\bibfield  {journal} {\bibinfo  {journal} {Phys.
  Rev. B}\ }\textbf {\bibinfo {volume} {79}},\ \bibinfo {pages} {174507}
  (\bibinfo {year} {2009})}\BibitemShut {NoStop}%
\bibitem [{Note6()}]{Note6}%
  \BibitemOpen
  \bibinfo {note} {This possibility was actually first mentioned by Leggett in
  1975 \cite {RevModPhys.47.331}, but no microscopic model exhibiting this
  behavior was known at the time. We thank Catherine Kallin for bringing this
  point to our attention.}\BibitemShut {Stop}%
\bibitem [{\citenamefont {Lederer}\ \emph {et~al.}(2015)\citenamefont
  {Lederer}, \citenamefont {Schattner}, \citenamefont {Berg},\ and\
  \citenamefont {Kivelson}}]{PhysRevLett.114.097001}%
  \BibitemOpen
  \bibfield  {author} {\bibinfo {author} {\bibfnamefont {S.}~\bibnamefont
  {Lederer}}, \bibinfo {author} {\bibfnamefont {Y.}~\bibnamefont {Schattner}},
  \bibinfo {author} {\bibfnamefont {E.}~\bibnamefont {Berg}}, \ and\ \bibinfo
  {author} {\bibfnamefont {S.~A.}\ \bibnamefont {Kivelson}},\ }\href {\doibase
  10.1103/PhysRevLett.114.097001} {\bibfield  {journal} {\bibinfo  {journal}
  {Phys. Rev. Lett.}\ }\textbf {\bibinfo {volume} {114}},\ \bibinfo {pages}
  {097001} (\bibinfo {year} {2015})}\BibitemShut {NoStop}%
\bibitem [{\citenamefont {Kozii}\ and\ \citenamefont
  {Fu}(2015)}]{PhysRevLett.115.207002}%
  \BibitemOpen
  \bibfield  {author} {\bibinfo {author} {\bibfnamefont {V.}~\bibnamefont
  {Kozii}}\ and\ \bibinfo {author} {\bibfnamefont {L.}~\bibnamefont {Fu}},\
  }\href {\doibase 10.1103/PhysRevLett.115.207002} {\bibfield  {journal}
  {\bibinfo  {journal} {Phys. Rev. Lett.}\ }\textbf {\bibinfo {volume} {115}},\
  \bibinfo {pages} {207002} (\bibinfo {year} {2015})}\BibitemShut {NoStop}%
\bibitem [{\citenamefont {Kang}\ and\ \citenamefont
  {Fernandes}(2016)}]{PhysRevLett.117.217003}%
  \BibitemOpen
  \bibfield  {author} {\bibinfo {author} {\bibfnamefont {J.}~\bibnamefont
  {Kang}}\ and\ \bibinfo {author} {\bibfnamefont {R.~M.}\ \bibnamefont
  {Fernandes}},\ }\href {\doibase 10.1103/PhysRevLett.117.217003} {\bibfield
  {journal} {\bibinfo  {journal} {Phys. Rev. Lett.}\ }\textbf {\bibinfo
  {volume} {117}},\ \bibinfo {pages} {217003} (\bibinfo {year}
  {2016})}\BibitemShut {NoStop}%
\bibitem [{\citenamefont {Wang}\ and\ \citenamefont
  {Chubukov}(2015)}]{PhysRevB.92.125108}%
  \BibitemOpen
  \bibfield  {author} {\bibinfo {author} {\bibfnamefont {Y.}~\bibnamefont
  {Wang}}\ and\ \bibinfo {author} {\bibfnamefont {A.~V.}\ \bibnamefont
  {Chubukov}},\ }\href {\doibase 10.1103/PhysRevB.92.125108} {\bibfield
  {journal} {\bibinfo  {journal} {Phys. Rev. B}\ }\textbf {\bibinfo {volume}
  {92}},\ \bibinfo {pages} {125108} (\bibinfo {year} {2015})}\BibitemShut
  {NoStop}%
\bibitem [{\citenamefont {Wang}\ \emph {et~al.}(2016)\citenamefont {Wang},
  \citenamefont {Cho}, \citenamefont {Hughes},\ and\ \citenamefont
  {Fradkin}}]{PhysRevB.93.134512}%
  \BibitemOpen
  \bibfield  {author} {\bibinfo {author} {\bibfnamefont {Y.}~\bibnamefont
  {Wang}}, \bibinfo {author} {\bibfnamefont {G.~Y.}\ \bibnamefont {Cho}},
  \bibinfo {author} {\bibfnamefont {T.~L.}\ \bibnamefont {Hughes}}, \ and\
  \bibinfo {author} {\bibfnamefont {E.}~\bibnamefont {Fradkin}},\ }\href
  {\doibase 10.1103/PhysRevB.93.134512} {\bibfield  {journal} {\bibinfo
  {journal} {Phys. Rev. B}\ }\textbf {\bibinfo {volume} {93}},\ \bibinfo
  {pages} {134512} (\bibinfo {year} {2016})}\BibitemShut {NoStop}%
\bibitem [{\citenamefont {Ruhman}\ \emph {et~al.}(2017)\citenamefont {Ruhman},
  \citenamefont {Kozii},\ and\ \citenamefont {Fu}}]{PhysRevLett.118.227001}%
  \BibitemOpen
  \bibfield  {author} {\bibinfo {author} {\bibfnamefont {J.}~\bibnamefont
  {Ruhman}}, \bibinfo {author} {\bibfnamefont {V.}~\bibnamefont {Kozii}}, \
  and\ \bibinfo {author} {\bibfnamefont {L.}~\bibnamefont {Fu}},\ }\href
  {\doibase 10.1103/PhysRevLett.118.227001} {\bibfield  {journal} {\bibinfo
  {journal} {Phys. Rev. Lett.}\ }\textbf {\bibinfo {volume} {118}},\ \bibinfo
  {pages} {227001} (\bibinfo {year} {2017})}\BibitemShut {NoStop}%
\bibitem [{\citenamefont {Kozii}\ \emph {et~al.}(2019)\citenamefont {Kozii},
  \citenamefont {Isobe}, \citenamefont {Venderbos},\ and\ \citenamefont
  {Fu}}]{PhysRevB.99.144507}%
  \BibitemOpen
  \bibfield  {author} {\bibinfo {author} {\bibfnamefont {V.}~\bibnamefont
  {Kozii}}, \bibinfo {author} {\bibfnamefont {H.}~\bibnamefont {Isobe}},
  \bibinfo {author} {\bibfnamefont {J.~W.~F.}\ \bibnamefont {Venderbos}}, \
  and\ \bibinfo {author} {\bibfnamefont {L.}~\bibnamefont {Fu}},\ }\href
  {\doibase 10.1103/PhysRevB.99.144507} {\bibfield  {journal} {\bibinfo
  {journal} {Phys. Rev. B}\ }\textbf {\bibinfo {volume} {99}},\ \bibinfo
  {pages} {144507} (\bibinfo {year} {2019})}\BibitemShut {NoStop}%
\bibitem [{\citenamefont {Zhao}\ \emph {et~al.}(2016)\citenamefont {Zhao},
  \citenamefont {Torchinsky}, \citenamefont {Chu}, \citenamefont {Ivanov},
  \citenamefont {Lifshitz}, \citenamefont {Flint}, \citenamefont {Qi},
  \citenamefont {Cao},\ and\ \citenamefont {Hsieh}}]{Zhao:2016aa}%
  \BibitemOpen
  \bibfield  {author} {\bibinfo {author} {\bibfnamefont {L.}~\bibnamefont
  {Zhao}}, \bibinfo {author} {\bibfnamefont {D.~H.}\ \bibnamefont
  {Torchinsky}}, \bibinfo {author} {\bibfnamefont {H.}~\bibnamefont {Chu}},
  \bibinfo {author} {\bibfnamefont {V.}~\bibnamefont {Ivanov}}, \bibinfo
  {author} {\bibfnamefont {R.}~\bibnamefont {Lifshitz}}, \bibinfo {author}
  {\bibfnamefont {R.}~\bibnamefont {Flint}}, \bibinfo {author} {\bibfnamefont
  {T.}~\bibnamefont {Qi}}, \bibinfo {author} {\bibfnamefont {G.}~\bibnamefont
  {Cao}}, \ and\ \bibinfo {author} {\bibfnamefont {D.}~\bibnamefont {Hsieh}},\
  }\href {\doibase 10.1038/nphys3517} {\bibfield  {journal} {\bibinfo
  {journal} {Nature Physics}\ }\textbf {\bibinfo {volume} {12}},\ \bibinfo
  {pages} {32} (\bibinfo {year} {2016})}\BibitemShut {NoStop}%
\bibitem [{\citenamefont {Zhao}\ \emph {et~al.}(2017)\citenamefont {Zhao},
  \citenamefont {Belvin}, \citenamefont {Liang}, \citenamefont {Bonn},
  \citenamefont {Hardy}, \citenamefont {Armitage},\ and\ \citenamefont
  {Hsieh}}]{Zhao:2017aa}%
  \BibitemOpen
  \bibfield  {author} {\bibinfo {author} {\bibfnamefont {L.}~\bibnamefont
  {Zhao}}, \bibinfo {author} {\bibfnamefont {C.~A.}\ \bibnamefont {Belvin}},
  \bibinfo {author} {\bibfnamefont {R.}~\bibnamefont {Liang}}, \bibinfo
  {author} {\bibfnamefont {D.~A.}\ \bibnamefont {Bonn}}, \bibinfo {author}
  {\bibfnamefont {W.~N.}\ \bibnamefont {Hardy}}, \bibinfo {author}
  {\bibfnamefont {N.~P.}\ \bibnamefont {Armitage}}, \ and\ \bibinfo {author}
  {\bibfnamefont {D.}~\bibnamefont {Hsieh}},\ }\href {\doibase
  10.1038/nphys3962} {\bibfield  {journal} {\bibinfo  {journal} {Nature
  Physics}\ }\textbf {\bibinfo {volume} {13}},\ \bibinfo {pages} {250}
  (\bibinfo {year} {2017})}\BibitemShut {NoStop}%
\bibitem [{\citenamefont {Xu}\ \emph {et~al.}(2019)\citenamefont {Xu},
  \citenamefont {Morimoto},\ and\ \citenamefont {Moore}}]{PhysRevB.100.220501}%
  \BibitemOpen
  \bibfield  {author} {\bibinfo {author} {\bibfnamefont {T.}~\bibnamefont
  {Xu}}, \bibinfo {author} {\bibfnamefont {T.}~\bibnamefont {Morimoto}}, \ and\
  \bibinfo {author} {\bibfnamefont {J.~E.}\ \bibnamefont {Moore}},\ }\href
  {\doibase 10.1103/PhysRevB.100.220501} {\bibfield  {journal} {\bibinfo
  {journal} {Phys. Rev. B}\ }\textbf {\bibinfo {volume} {100}},\ \bibinfo
  {pages} {220501} (\bibinfo {year} {2019})}\BibitemShut {NoStop}%
\bibitem [{\citenamefont {Kashiwaya}\ \emph {et~al.}(2011)\citenamefont
  {Kashiwaya}, \citenamefont {Kashiwaya}, \citenamefont {Kambara},
  \citenamefont {Furuta}, \citenamefont {Yaguchi}, \citenamefont {Tanaka},\
  and\ \citenamefont {Maeno}}]{KashiwayaEA11}%
  \BibitemOpen
  \bibfield  {author} {\bibinfo {author} {\bibfnamefont {S.}~\bibnamefont
  {Kashiwaya}}, \bibinfo {author} {\bibfnamefont {H.}~\bibnamefont
  {Kashiwaya}}, \bibinfo {author} {\bibfnamefont {H.}~\bibnamefont {Kambara}},
  \bibinfo {author} {\bibfnamefont {T.}~\bibnamefont {Furuta}}, \bibinfo
  {author} {\bibfnamefont {H.}~\bibnamefont {Yaguchi}}, \bibinfo {author}
  {\bibfnamefont {Y.}~\bibnamefont {Tanaka}}, \ and\ \bibinfo {author}
  {\bibfnamefont {Y.}~\bibnamefont {Maeno}},\ }\href {\doibase
  10.1103/PhysRevLett.107.077003} {\bibfield  {journal} {\bibinfo  {journal}
  {Phys. Rev. Lett.}\ }\textbf {\bibinfo {volume} {107}},\ \bibinfo {pages}
  {077003} (\bibinfo {year} {2011})}\BibitemShut {NoStop}%
\bibitem [{\citenamefont {Chronister}\ \emph {et~al.}(2020)\citenamefont
  {Chronister}, \citenamefont {Pustogow}, \citenamefont {Kikugawa},
  \citenamefont {Sokolov}, \citenamefont {Jerzembeck}, \citenamefont {{Hicks}},
  \citenamefont {{Mackenzie}}, \citenamefont {Bauer},\ and\ \citenamefont
  {Brown}}]{Chronister2020}%
  \BibitemOpen
  \bibfield  {author} {\bibinfo {author} {\bibfnamefont {A.}~\bibnamefont
  {Chronister}}, \bibinfo {author} {\bibfnamefont {A.}~\bibnamefont
  {Pustogow}}, \bibinfo {author} {\bibfnamefont {N.}~\bibnamefont {Kikugawa}},
  \bibinfo {author} {\bibfnamefont {D.~A.}\ \bibnamefont {Sokolov}}, \bibinfo
  {author} {\bibfnamefont {F.}~\bibnamefont {Jerzembeck}}, \bibinfo {author}
  {\bibfnamefont {C.~W.}\ \bibnamefont {{Hicks}}}, \bibinfo {author}
  {\bibfnamefont {A.~P.}\ \bibnamefont {{Mackenzie}}}, \bibinfo {author}
  {\bibfnamefont {E.~D.}\ \bibnamefont {Bauer}}, \ and\ \bibinfo {author}
  {\bibfnamefont {S.~E.}\ \bibnamefont {Brown}},\ }\href@noop {} {\bibfield
  {journal} {\bibinfo  {journal} {arXiv}\ } (\bibinfo {year}
  {2020})}\BibitemShut {NoStop}%
\end{thebibliography}%

\appendix
\onecolumngrid
\newpage\clearpage

\section{Analytic solution of the weak coupling equation}
Within a weak coupling analysis of the superconducting instability, we have to solve the following equation:
\bea
\frac1{(2\pi)^2} \int \frac{d\hat{k}_2}{|v(\hat{k}_2)|} V(\hat{k}_1 - \hat{k}_2) \Delta(\hat{k}_2) = \lambda \ \Delta(\hat{k}_1)
\label{EigenEqApp}
\eea
where
\bea
V_e(\hat{k}_1 - \hat{k}_2) &= U + U^2 \chi(\hat{k}_1 - \hat{k}_2) \\
V_o(\hat{k}_1 - \hat{k}_2) &= - U^2 \chi(\hat{k}_1 - \hat{k}_2)
\label{effectiveinteraction} 
\eea
are the effective interactions in the even and odd sector, where $\chi$ is the Lindhard susceptibility, and where $\hat{k}_1, \hat{k}_2$ live on the Fermi surface.
In this appendix, we will provide an analytic solution that is valid in the limit of $t_y / t_x \rightarrow 0$.

In this limit, the Fermi surfaces are given by two sheets at $k_x= \pm k_F(k_y)$, with
\bea
k_F(k_y) = k_F + \frac{2 t_y \cos(k_y)}{v_F} + \mathcal{O}(t_y^2)
\eea
where $v_F = 2 t_x \sin(k_F)$ and $k_F = \mathrm{arccos}(- \mu / 2 t_x)$.

\subsection{Lindhard susceptibility}

Since the Fermi surface is given by two separate sheets at $k_x \simeq k_F$, we only need the value of $\chi(q_x,q_y)$ in two regimes: for $q_x \simeq 0$ (for intra-sheet scattering), and for $q_x \simeq 2 k_F$ (for inter-sheet scattering).
For intra-sheet scattering, one easily finds that
\bea
\chi(q_x \simeq 0 , q_y) = \rho + \mathcal{O}(t_y).
\eea
with $\rho$ the density of states at the Fermi level  in the vanishing $t_y$ limit.
We can therefore forget about intra-sheet scattering since this constant term will only give a contribution in the trivial $s$-wave channel.

The inter-sheet case is more interesting: we will find that
\bea
\chi(q_x \simeq 2 k_F , q_y) &= C(t_y) + f(q_y) + \mathcal{O}(t_y) 
\eea
where $C(t_y)$ is an unimportant constant since it will only give a contribution in the trivial $s$-wave channel, and where $f(q_y)$ is a non-trivial function that is independent of $t_y$ and that will need to be diagonalized in order to solve the problem at hand.

\begin{figure*}[t!]
\center
  \includegraphics[width=6cm]{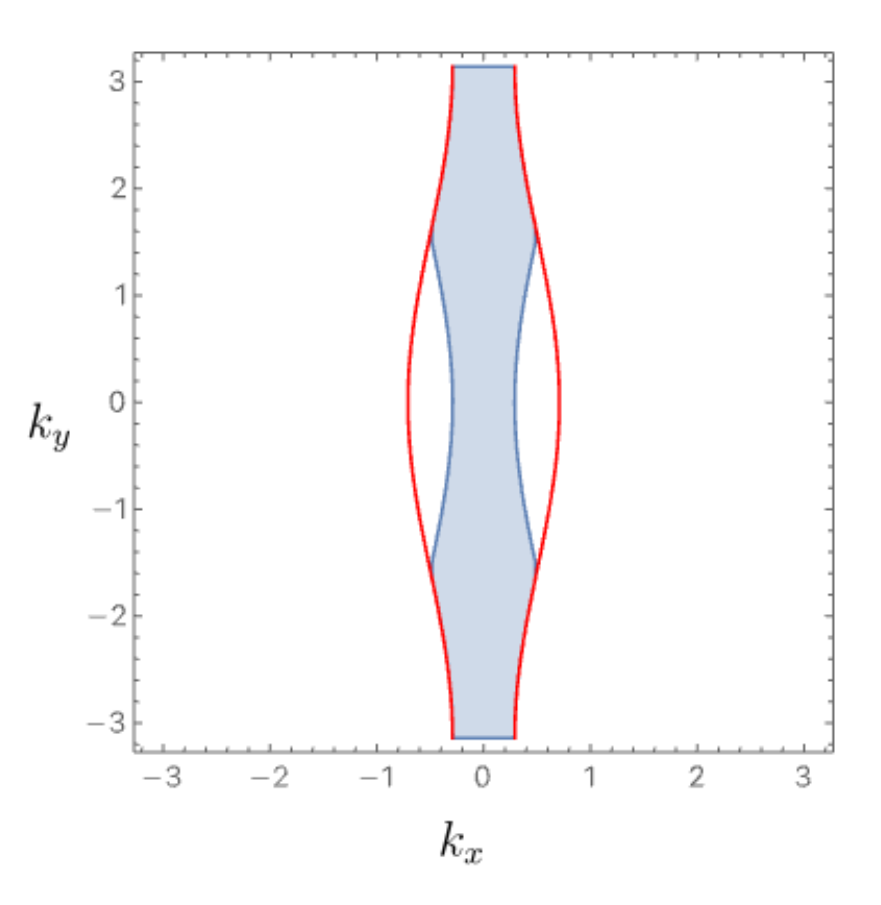}
  \caption{The blue region shows a typical example of the zone 2 defined in Eq.~\ref{Eq:Zone2}. The Fermi surface is shown in red. Parameters are $k_{F}=0.5$, $t_y/t_x=0.1$, $q_y=0$, and $q_x=1$. }
  \label{Zone2}
\end{figure*}

As a reminder, the susceptibility is defined as:
\bea
\chi(\bq) = -\frac1{(2\pi)^2} \int d\bp \frac{n(\xi(\bp)) - n(\xi(\bp+\bq))}{\xi(\bp) - \xi(\bp+\bq)}
\eea
The numerator is non-zero in two disjoints regions, one for which $\xi(\bp) > 0$ and $\xi(\bp+\bq))<0$ (zone 1) and one for which $\xi(\bp) < 0$ and $\xi(\bp+\bq))>0$ (zone 2).
Since these two zones give the same contribution to the integral, we will only focus on zone 2.
For a given $k_y$, the zone limits for zone 2 are $k_{x,start} \leq k_x \leq k_{x,end}$ with
\bea
k_{x,start}(k_y) &= \mathrm{max}(k_F(k_y+q_y)-q_x, -k_F(k_y)) \\
k_{x,end}(k_y) &= \mathrm{min}(-k_F(k_y+q_y)+q_x, k_F(k_y)).
\label{Eq:Zone2}
\eea
A typical example of zone 2 is shown in Fig.~\ref{Zone2}.

We are now interested in the locus of points $\tilde{k}_x$ where the denominator vanishes (i.e. where $\xi(\bp) - \xi(\bp+\bq) = 0$) since the integrand will be peaked there. It is given, to leading order in $t_y$, by
\bea
\tilde{k}_x(k_y) = -k_F + \frac12 ( k_F(k_y+q_y) - k_F(k_y) - (q_x - 2k_F))
\eea
It will be useful to define 
$k_y^*$ as
\bea
k_F(k_y^* + q_y) = -k_F(k_y^*) + q_x.
\eea
In other words, $k_y^*(q_x,q_y)$ is the value of $k_y$ on the left branch such that $k+q$ sits exactly on the right branch at $k_y + q_y$.
With this parametrization, we find $q_x = k_F(k_y^*+q_y) + k_F(k_y^*)$.

Now, we can expand the denominator $D(p_x,p_y) \equiv \xi(\bp) - \xi(\bp+\bq) $ linearly along the $k_x$ direction:
\bea
D(p_x,p_y) \simeq (p_x - \tilde{k}_x) \ (\partial_{p_x}D)|_{\tilde{k}_x} + \mathcal{O}((p_x - \tilde{k}_x)^2)
\eea
To leading order in $t_y$, we find $(\partial_{p_x}D)|_{\tilde{k}_x} = -2 v_F$, thus $D(p_x,p_y) \simeq (p_x - \tilde{k_x}) 2 v_F$.

Now, the integral becomes:
\bea
\chi(\bq)  &= -\frac2{(2\pi)^2} \int dp_y \int_{k_{x,start}(p_y)}^{k_{x,end}(p_y)} dp_x \frac{1}{-2 v_F (p_x - \tilde{k_x})} \\
&= \frac2{2 v_F (2\pi)^2} \int dp_y (\log(k_{x,end}(p_y)-\tilde{k}_x(p_y)) - \log(k_{x,start}(p_y) -\tilde{k}_x(p_y)) ) \\
&\simeq -\frac2{2 v_F (2\pi)^2} \int dp_y   \log(k_{x,start}(p_y) -\tilde{k}_x(p_y)) 
\eea
where in the last line, we used the fact that, in the small $t_y$ limit, $k_{x,start}(p_y) -\tilde{k}_x(p_y)$ goes to zero, while $k_{x,end}(p_y) -\tilde{k}_x(p_y)$ is finite.
We also find that
\bea
k_{x,start}(p_y)-\tilde{k}_x(p_y) &= |\tilde{k}_x - (-k_F(p_y))| \\
&= \frac12 |k_F(p_y) + k_F(p_y+q_y) - k_F(k_y^* + q_y) - k_F(k_y^*)| \\
&= \frac{2 t_y}{v_F} \frac12 |\cos(p_y) + \cos(p_y+q_y) - \cos(k_y^* + q_y) - \cos(k_y^*)| + \mathcal{O}(t_y^2)
\eea 
which finally leads to
\bea
\chi(q) &= -2\frac1{2 v_F (2\pi)^2} \int dp_y   \log\left(\frac{2 t_y}{v_F} \frac12 |\cos(p_y) + \cos(p_y+q_y) - \cos(k_y^* + q_y) - \cos(k_y^*)|\right).  
\eea

After some algebra, we find the simple relation:
\bea
\chi(\hat{k}_2 - \hat{k}_1) &=  \chi'_0 -\frac1{2 v_F (2\pi)}   \log\left(  \cos(k_{y,2}-k_{y,1})+1 \right)
\label{chiAnalytic}
\eea 
 with $\chi_0' = \frac1{2 v_F (2\pi)}   \log(\sqrt{2} v_F / t_y)$ an inconsequential constant since it will only give a repulsive contribution in the $m=0$ channel (see below).

  \subsection{Diagonalization}
Starting from the initial eigenproblem (Eq.\ref{EigenEqApp}), we can make a further set of approximations which are valid to leading order in $t_y$. We can omit constant terms in the effective interaction, since they will only contribute to the $m=0$ sector, which is always repulsive. This includes the $U$ term in Eq.~\ref{effectiveinteraction}, the intra-sheet scattering (i.e. when $\hat{k}_1$ and $\hat{k}_2$ are on the same FS sheet), and the $\chi'_0$ term in Eq.\ref{chiAnalytic}.
Finally, to leading order, we can take the Fermi velocity to be constant: $v(k) = v_F$.
After all these approximations, the even and odd-parity sector eigenproblems both simplify to the same equation:
 \bea
\frac1{(2\pi)^2 v_F} s_y \int_{-\pi}^\pi dk_{y,2} \ \chi(k_{y,1}-k_{y,2}) \Delta(k_{y,2}) = \lambda \ \Delta(k_{y,1})
\label{FinalEigProb}
\eea
where $\chi(k_{y,1}-k_{y,2})$ is given in Eq.~\ref{chiAnalytic}, and where $s_y$ is the sign change of $\Delta$ under the $y \rightarrow -y$ mirror symmetry.
 
 Since $\chi$ only depends on $k_{y,2} - k_{y,1}$, we can always diagonalize Eq.~\ref{FinalEigProb} with Fourier series, leading to four sets of eigenvectors:
 \bea
 \Delta_{m,1,1}(k_x,k_y) &= \cos(m k_y) \\
 \Delta_{m,-1,1}(k_x,k_y) &= \cos(m k_y) \mathrm{sign}(k_x) \\
  \Delta_{m,1,-1}(k_x,k_y) &= \sin(m k_y) \\
    \Delta_{m,-1,-1}(k_x,k_y) &= \sin(m k_y) \mathrm{sign}(k_x) \\
 \eea
 for $m \geq 1$ (it is easy to check that the $m=0$ states are repulsive). Using the relation
 \bea
\frac1{2\pi} \int_{-\pi}^\pi dk_{y}     \log\left(  \cos(k_{y})+1 \right) \cos(m k_y) = \frac{(-1)^{m+1}}m
\eea
valid for $m \geq 1$, one finds all the negative eigenvalues:
\bea
 \lambda_{m} = -  \frac{U^2}{2 (2\pi)^2 v_F^2} \frac1m \\
 \eea
 for $m \geq 1$.
 Each of these eigenvalues is doubly degenerate. 
 For odd $m$, the eigenvectors are given by
\bea
\Delta_{m,e} &= \Delta_{m,1,1}(k_x,k_y) = \cos(m k_y) \\
\Delta_{m,o} &=   \Delta_{m,-1,1}(k_x,k_y) = \cos(m k_y) \mathrm{sign}(k_x).
\eea
For even $m$, we find
\bea
\Delta_{m,e} &=     \Delta_{m,-1,-1}(k_x,k_y)  = \sin(m k_y)  \mathrm{sign}(k_x)\\
\Delta_{m,o} &= \Delta_{m,1,-1}(k_x,k_y) = \sin(m k_y).
\eea

\section{Ginzburg-Landau analysis}
\label{App:GL}
We consider the following mixed parity order parameter:
 \bea
\mathbf{\Delta}(\bk) \equiv \Delta_{\up\down}(\bk) = \psi_e \Delta_e(\bk) + i \psi_o \Delta_o(\bk)
\eea
with $\psi_e$ and $\psi_o$ real parameters, leading to $|\mathbf{\Delta}(\bk)|^2=|\psi_e \Delta_e(\bk)|^2 + |\psi_o \Delta_o(\bk)|^2$.
The free energy is given by
\bea
F = -a_e \psi_e^2 - a_o \psi_o^2 + b (\psi_e^2 + \psi_o^2)^2 + b' (\psi_e^2 - \psi_o^2)^2
\eea
with $a_e(T) = a_0 (T_{c,e} -T)$, $a_o(T) = a_0 (T_{c,o} -T)$, where $T_{c,e}$ and $T_{c,o}$ are the critical temperatures for each component when considered in isolation.
  
  The first transition occurs at $T_{c,e}$ and the second transition occurs at $T^*$ given by
  \bea
T^* = T_{c,o} - (T_{c,e} - T_{c,o}) \frac{b-b'}{2b'}.
\eea
  The solution for the order parameter is
  \bea
\left(\psi_{e,0}^2 = \frac{a_e}{2(b+b')}, \psi_{t,0}^2=0 \right)
\eea
for $T^* < T < T_{c,e}$ and
 \bea
\left( \psi_{e,1}^2 = \frac12 \left( \frac{a_e + a_o}{4 b} + \frac{a_e - a_o}{4 b'} \right),  \psi_{o,1}^2 = \frac12 \left( \frac{a_e + a_o}{4 b} - \frac{a_e - a_o}{4 b'} \right) \right)
\eea
for $T < T^*$.

  \subsection{Specific heat discontinuity}
  \label{App:CJump}
 The specific heat discontinuity at a temperature $T$ is given by \cite{Sigrist05}:
\bea
\Delta C = N_0 \frac{\beta}4 \int d\xi \left\langle \frac1{\cosh(\beta E/2)^2} \left( -\left(\frac{\partial |\mathbf{\Delta}(\bk)|^2}{\partial T} \right)_{T_-} + \left(\frac{\partial |\mathbf{\Delta}(\bk)|^2}{\partial T} \right)_{T_+} \right)\right\rangle_{FS}
\eea
with
\bea
\left\langle f(k) \right\rangle_{FS} &= \frac1{N_0} \frac1{(2\pi)^D} \int_{FS} d\hat{k} \frac1{v(k)} f(\hat{k}) \\
N_0 &= \frac1{(2\pi)^D} \int_{FS} d\hat{k} \frac1{v(k)}
\eea
 and where $T_+$ and $T_-$ are approaching $T$ from above and below, respectively.
 
At the first transition, this leads to
\bea
\Delta C_{T_{c,e}} &= N_0 \frac{\beta}4 \int d\xi  \frac1{\cosh(\beta E/2)^2} \left\langle - \left( \frac{\partial |\mathbf{\Delta}(\bk)|^2}{\partial T} \right)_{T_-} \right\rangle_{FS} \\
&= N_0 \frac{\beta}4 \int d\xi  \frac1{\cosh(\beta \xi/2)^2} \left\langle |\Delta_e(\bk)|^2 \right\rangle_{FS} \frac{a_0}{2(b+b')} \\
&= N_0 \left\langle |\Delta_e(\bk)|^2 \right\rangle_{FS} \frac{a_0}{2(b+b')} 
\eea
where we used
\bea
\int d\xi \frac1{\cosh(\beta \xi/2)^2} = \frac{4}{\beta}.
\eea
 
 At the second transition, we find
 \bea
\Delta C_{T^*} &= N_0 \frac{\beta}4 \int d\xi \left\langle \frac1{\cosh(\beta E/2)^2} \left( -\left(\frac{\partial |\mathbf{\Delta}(\bk)|^2}{\partial T} \right)_{T_-} + \left(\frac{\partial |\mathbf{\Delta}(\bk)|^2}{\partial T} \right)_{T_+} \right)\right\rangle_{FS} \\
&= N_0 \frac{\beta}4 \int d\xi \left\langle \frac1{\cosh(\beta E/2)^2}  |\Delta(\bk)|^2 \right\rangle_{FS}  \left(  \frac{a_0}{2b} - \frac{a_0}{2(b+b')} \right) \\
&= N_0   \left\langle Y_{T^*}(\bk)  |\Delta(\bk)|^2 \right\rangle_{FS}  \left(  \frac{a_0}{2b} - \frac{a_0}{2(b+b')} \right)
\eea
 where we made the approximation that $|\Delta_e(\bk)|^2 = |\Delta_o(\bk)|^2 \equiv |\Delta(\bk)|^2$.
 
 Eq.~\ref{SpecHeat} of the main text is obtained by taking the ratio of the two specific heat discontinuities, combined with the approximation $|\Delta_e(\bk)|^2 = |\Delta_o(\bk)|^2 \equiv |\Delta(\bk)|^2$.
 
 \section{Spin susceptibility}
 \label{App:SpinSus}
 In this section, we calculate the spin susceptibility of a mixed singlet-triplet superconductor when the $\vec{H}$ field is perpendicular to $\vec{d}$.
Without loss of generality, we choose $\vec{H} \parallel \hat{z}$ and $\vec{d} = -i \psi_o \Delta_o(\bk)  \hat{y}$.
 For a given $\bk$ value, the BCS Hamiltonian reads \cite{Sigrist05}
\bea
H = 
\begin{pmatrix}
c^\dagger_\uparrow(\bk) & c^\dagger_\downarrow(\bk) & c_\uparrow(\bk) & c_\downarrow(\bk)
\end{pmatrix}	
\begin{pmatrix}
 \xi(\bk)+h & 0 & \psi_o \Delta_o(\bk) & \psi_e\Delta_e(\bk) \\
 0  & \xi(\bk) - h & -  \psi_e\Delta_e(\bk) &  \psi_o\Delta_o(\bk)\\
  \psi_o\Delta_o(\bk) & - \psi_e\Delta_e(\bk) & -(\xi(\bk)+h)& 0 \\
   \psi_e\Delta_e(\bk) & \psi_o\Delta_o(\bk) & 0& -(\xi(\bk)-h)
\end{pmatrix}
\begin{pmatrix}
c_\uparrow(\bk) \\ c_\downarrow(\bk) \\ c^\dagger_\uparrow(\bk) \\ c^\dagger_\downarrow(\bk)
\end{pmatrix}	
\eea
where $h$ is the Zeeman splitting.
The resulting magnetization can be calculated analytically by performing a Bogolyubov transformation, and the magnetic susceptibility is read off from the linear-in-$h$ term.

Up to an overall multiplicative constant, the magnetic susceptibility is found to be 
\bea
\chi &= \sum_{\bk} \frac{\beta}{4} \frac1{\cosh(\frac12 \beta E)^2} \left( 1 - \frac{|\psi_o \Delta_o(\bk)|^2}{E^2} \left(1 - \frac{\sinh(\beta E)}{\beta E}\right) \right) \\
&= \frac1{(2\pi)^D} \int d\xi \int d\hat{k} \frac1{v(k)} \frac{\beta}{4} \frac1{\cosh(\frac12 \beta E)^2} \left( 1 - \frac{|\psi_o \Delta_o(\bk)|^2}{E^2} \left(1 - \frac{\sinh(\beta E)}{\beta E}\right) \right) \\
&= N_0 \avg{Y_T(k)} - N_0 \avg{\frac{\beta}{4} \int d\xi\frac1{\cosh(\frac12 \beta E)^2}  \frac{|\psi_o \Delta_o(\bk)|^2}{E^2} \left(1 - \frac{\sinh(\beta E)}{\beta E}\right) } \\
&= N_0 \avg{Y_T(k)} + N_0 \avg{\frac{|\psi_o \Delta_o(\bk)|^2}{|\mathbf{\Delta(\bk)}|^2}  (1-Y_T(k))}
\eea
with $|\mathbf{\Delta(\bk)}|^2 = |\psi_o \Delta_o(\bk)|^2 + |\psi_e \Delta_e(\bk)|^2$.

 Since the normal state susceptibility is given by $N_0$ in our units, one recovers Eq.~\ref{Eq:MagSus} from the main text.

\end{document}